\documentclass[preprint,aps,pra,showpacs,floatfix,superscriptaddress]{revtex4-1}
\usepackage{graphicx}
\usepackage{feynmp}
\usepackage{amsmath,amsfonts,amssymb,verbatim,float}
\usepackage{epsf}
\usepackage{bm}
\usepackage{bbm}
\usepackage[usenames]{color}	%for colours
\usepackage{colortbl}
\usepackage{hyperref} % for links in the bibliography
\usepackage{indentfirst}
\usepackage{siunitx} % vyravnivanie po tichke
\usepackage{calrsfs}
\usepackage{calrsfs}
\usepackage{ulem} % for remarks
\usepackage{subfigure}
\newcommand{\br}{{\bf r}}

\newcommand{\balpha}{\boldsymbol{\alpha}}

\graphicspath{{graph/}}
\newcommand{\be}{\begin{eqnarray}}
\newcommand{\ee}{\end{eqnarray}}
\newcommand{\la}{\langle}
\newcommand{\ra}{\rangle}

\newcommand{\veps}{\varepsilon}

\newcommand{\bnabla}{\bm{\nabla}}

\newcommand{\rmd}{{\rm d}}

\newcommand{\bfr}{{\bf r}}

\newcommand{\ket}[1]{|#1\rangle}
\newcommand{\braket}[2]{\langle#1|#2\rangle}
\newcommand{\matrixel}[3]{\langle #1 | #2 | #3 \rangle}

\newcommand{\tomega}{\tilde\omega}
\newcommand{\gdirac}{g_{\rm D}}
\newcommand{\dgint}{\Delta g_{\rm int}}
\newcommand{\dgqed}{\Delta g_{\rm QED}}

\newcommand{\dgnuc}{\Delta g_{\rm nuc}}
\begin{document}
\thispagestyle{empty}
\title{
Electron correlation effects on the $g$ factor of lithiumlike ions
}
\author{D. V. Zinenko}
\affiliation{Department of Physics, St. Petersburg State University, Universitetskaya nab. 7/9, 199034 St. Petersburg, Russia}
\author{D. A. Glazov}
\affiliation{Department of Physics, St. Petersburg State University, Universitetskaya nab. 7/9, 199034 St. Petersburg, Russia}
\author{V. P. Kosheleva}
\affiliation{Max Planck Institute for the Structure and Dynamics of Matter and Center for Free-Electron Laser Science, Hamburg 22761, Germany}
\author{A. V. Volotka}
\affiliation{School of Physics and Engineering, ITMO University, Kronverkskiy pr. 49, 197101 St. Petersburg, Russia}
\author{S. Fritzsche}
\affiliation{Theoretisch-Physikalisches Institut, Friedrich-Schiller-Universität Jena, Max-Wien-Platz 1, 07743 Jena, Germany}
%
%
%======================       ABSTRACT         =========================
\begin{abstract}
%The rigorous QED evaluation of the one- and two-photon exchange corrections to the $g$ factor of Li-like ions is presented for the wide range of nuclear charge number $Z= 14 - 82$.
%
We present the systematic QED treatment of the electron correlation effects on the $g$ factor of lithiumlike ions for the wide range of nuclear charge number $Z= 14$ -- $82$.
The one- and two-photon exchange corrections are evaluated rigorously within the QED formalism. 
The electron-correlation contributions of the third and higher orders are accounted for within the Breit approximation employing the recursive perturbation theory.
%
%The interelectronic-interaction contributions of the third and higher orders are taken into account in the framework of the Breit approximation employing the recursive perturbation theory. 
%
The calculations are performed in the framework of the extended Furry picture, i.e., with inclusion of the effective local screening potential in the zeroth-order approximation.
In comparison to the previous theoretical calculations, the accuracy of the interelectronic-interaction contributions to the bound electron $g$ factor in lithiumlike ions is substantially improved.
\end{abstract}
%======================    END ABSTRACT       ==========================
%

%%%%%%%%%%%%%%%%%%%%%%%%%%%%%%%%%%%%%%%%%%%%%%%%%%%%%%%%%%%%%%%%%%%%%%%%
\maketitle
%%%%%%%%%%%%%%%%%%%%%%%%%%%%%%%%%%%%%%%%%%%%%%%%%%%%%%%%%%%%%%%%%%%%%%%%
%
% ======================== INTRODUCTION ================================
%
\section{Introduction}
\label{sec:intro}
Over the past decades, the bound-electron $g$ factor remains a subject of intense theoretical and experimental studies.
Nowadays, the $g$ factor of H-like ions is measured with a relative accuracy of up to few parts in $10^{11}$  \cite{haeffner:2000:5308,verdu:2004:093002,sturm:2011:023002,sturm:2013:R030501,sturm:2014:467, sailer:2022:479}. 
These measurements combined with the theoretical studies \cite{persson:1997:R2499, blundell:1997:1857, beier:2000:79, karshenboim:2000:380, karshenboim:2001:81, glazov:2002:408, shabaev:2002:091801, nefiodov:2002:081802, yerokhin:2002:143001, pachucki:2005:022108, jentschura:2009:044501, yerokhin:2013:042502, yerokhin:2013:245002} have led to the most accurate up-to-date value of the electron mass \cite{sturm:2014:467}.
Present experimental techniques also allow for the $g$-factor measurements in few-electron ions \cite{wagner:2013:033003, lindenfels:2013:023412, koehler:2016:10246, glazov:2019:173001, arapoglou:2019:253001, egl:19:prl, micke:20:n} with the accuracy comparable to that for H-like ions.
%
%In particular, the recent experiments for Li-like ions report results with eleven significant digits~\cite{koehler:2016:10246, glazov:2019:173001}, thus reaching 

High-precision measurements of the $g$ factor of highly charged ions provide various opportunities to probe the non-trivial QED effects in strong electromagnetic fields, determine fundamental constants and nuclear properties, and strengthen the limits on the hypothetical physics beyond the Standard Model~\cite{yerokhin:2011:043004, volotka:2013:636, shabaev:2015:031205, indelicato:2019:232001, debierre:2020:135527, shabaev:2022:043001, debierre:2022:062801}.
For example, the measurement of the $g$-factor isotope shift with lithiumlike calcium ions \cite{koehler:2016:10246} has opened a possibility to test the relativistic nuclear recoil theory in the presence of magnetic field and paved the way to probe bound-state QED effects beyond the Furry picture in the strong-field regime \cite{shabaev:2017:263001, malyshev:2017:731, shabaev:2018:032512}.
The high-precision bound-electron $g$-factor experiments combined with theoretical studies are expected to provide an independent determination of the fine structure constant $\alpha$ \cite{shabaev:2006:253002, yerokhin:2016:100801}. 
While the $g$-factor calculations are progressing further \cite{czarnecki:2016:060501, yerokhin:2017:060501, czarnecki:2018:043203, sikora:2020:012002, oreshkina:2020:032511, czarnecki:2020:050801, debierre:2021:L030802}, the accuracy of theoretical values is ultimately limited by the finite-nuclear-size effect. To overcome this problem, it was proposed to use the so-called specific difference of the $g$ factors for two charge states of one isotope \cite{shabaev:2002:062104, shabaev:2006:253002, volotka:2014:023002, yerokhin:2016:100801, yerokhin:2016:022502, malyshev:2017:731}. It was demonstrated that the theoretical uncertainty of the specific difference can be made several orders of magnitude smaller than that of the individual $g$-factor values. Therefore, it is very important to consider not only hydrogenlike but also lithiumlike and boronlike ions.

The first experiments with lithiumlike ions were carried out for silicon \cite{wagner:2013:033003} and calcium \cite{koehler:2016:10246} with an uncertainty of $10^{-9}$. Recently, the experimental value of the bound-electron $g$ factor in $^{28}$Si$^{11+}$ was improved by a factor of 15 \cite{glazov:2019:173001}, and this is currently the most accurate value for few-electron ions.
The theoretical value presented in Ref.~\cite{glazov:2019:173001} was 2 times more accurate than the previous one \cite{volotka:2014:253004, yerokhin:2017:062511},
% At that time, this was the most accurate theoretical calculation, which relied on the. 
% includes many-electron QED terms obtained within the rigorous approach  and higher-order contributions was , 
while the deviation from the experiment was $1.7\sigma$. Later, Yerokhin {\it et al.} undertook an independent evaluation of screened QED diagrams and obtained a new theoretical value for silicon \cite{yerokhin:2020:022815} with smaller uncertainty and shifted farther from the experiment: $5.2\sigma$ deviation as a result. Then an independent evaluation of the two-photon-exchange diagrams was carried out to provide yet new results for lithiumlike silicon (with $3.1\sigma$ deviation) and calcium (with $4.2\sigma$ deviation)~\cite{yerokhin:2021:022814}. Recently, we have thoroughly investigated the behaviour of many-electron QED contributions with various effective screening potentials~\cite{kosheleva:2022:103001}, thus confirmed the results of Ref.~\cite{glazov:2019:173001} and reassured the agreement between theory and experiment.
% found that the theoretical value for $^{28}$Si$^{11+}$ agrees with the previous theoretical result and does not significantly differ from experiment. Meanwhile for $^{40}$Ca$^{17+}$ we obtained an excellent agreement with the experiment.

In the present work, we provide detailed investigation of the electronic structure contributions and extend our calculations to a wide range of the nuclear charge number $Z$. The leading-order interelectronic-interaction terms corresponding to the one- and two-photon-exchange diagrams nowadays are calculated rigorously, i.e., to all orders in $\alpha Z$~\cite{volotka:2012:073001, wagner:2013:033003, volotka:2014:253004, yerokhin:2021:022814, kosheleva:2022:103001}. The contributions of the third and higher orders are taken into account approximately, to the leading orders in $\alpha Z$. This can be accomplished within different methods, which can yield slightly different results due to the incomplete treatment of the higher orders in $\alpha Z$.
%in the NRQED approach using the accurate variational wave functions in the Hyleraas basis \cite{yerokhin:2017:062511}, by employing the configuration interaction method \cite{bratsev:1977:2655,glazov:2004:062104}, or within the recursive perturbation theory \cite{glazov:2017:46,glazov:2019:173001} in the Breit approximation.
%
We use the Dirac-Coulomb-Breit (DCB) Hamiltonian and include the contribution of the negative-energy states~\cite{glazov:2004:062104}. In Refs.~\cite{glazov:2004:062104, glazov:2005:55, volotka:2014:253004} the DCB equation was solved within the configuration interaction method \cite{bratsev:1977:2655}. Later, in Refs.~\cite{glazov:2019:173001, kosheleva:2022:103001} the recursive perturbation theory was applied and proved to yield better accuracy.
Moreover, we use the extended Furry picture, i.e., an effective local screening potential is included in the zeroth-order approximation along with the nuclear potential. 
%
%The first two terms are evaluated (one- and two- photon exchange) rigorously within QED formalism.
%
%The higher-order interelectronic-interaction contributions is calculated by means of the recursive perturbation theory.
%
The extended Furry picture provides partial account for the interelectronic interaction already in the zeroth order. As a result, there is a significant reduction of the perturbation theory terms in comparison to the case of the Coulomb potential. In particular, this leads to smaller uncertainty due to unknown non-trivial QED part of the third and higher-order contributions.
The calculations are carried out with different screening potentials, namely, core-Hartree (CH), Kohn-Sham (KS), Dirac-Hartree (DH), and Dirac-Slater (DS), with and without the Latter correction. We show that the influence of this correction on the final result is insignificant, while the accuracy of calculations without it is better by an order of magnitude. In the QED calculations of the first and second orders, two different gauges of the photon propagator, Coulomb and Feynman, are considered and the gauge invariance is demonstrated, which serves as an additional test of the correctness of our calculations.
As a result, we substantially improve the accuracy of the electronic-structure contribution to the $g$ factor of lithiumlike ions through a wide range of the nuclear charge number $Z= 14$ -- $82$.

The paper is organized as follows. In Sec. II, the basic formulas for the bound-electron $g$ factor in few-electron ions are given. In Sec. III, we present the rigorous theoretical description of the interelectronic-interaction corrections of the first (III A), second (III B), and higher orders (III C). Finally, in Sec. IV, we report the obtained numerical results.

Relativistic units ($\hbar = 1$, $c = 1$, $m_e = 1$) and the Heaviside charge unit [$\alpha = e^2/(4\pi)$, $e<0$] are used throughout the paper.
%
% ======================== BASIC FORMALISM =============================
%
\section{Basic formulas}
\label{sec:basic}
The interaction of the bound electron with the external magnetic field $\boldsymbol{B}$ is represented by the operator,
\be
V_{\rm m} = -e\bm{\alpha}\cdot \bm{A(r)}=-\frac{e}{2}BU\,,
\label{eq:V_magn} 
\ee
where, without loss of generality, $B$ is assumed to be aligned in $z$-direction, $U = [\bfr\times \balpha]_z$ and $\balpha$ is the Dirac-matrix vector.
Weak magnetic field induces the linear energy level shift,
\be
  \Delta E  = -\dfrac{e}{2} g m_j B \,,
  \label{eq:e_shift} 
\ee 
where $g$ is the electronic $g$ factor and $m_j$ is the $z$-projection of the total angular momentum $j$.
In the case of one electron over the closed shells and a spinless nucleus, $m_j$ is determined by the valence electron state $\ket{a} = \ket{j_a m_a}$, with the angular momentum $j_a$ and its projection $m_a$. In the ground $(1s)^2\,2s$ state of a lithiumlike atom, this is just the $2s$ state.
%
%where index $i$ refers to the $i$th electron of the atom.
%
%
%The lowest-order $g$ factor in the one-electron Dirac approximation for the case of pointlike Coulomb potential of the nucleus is known analytically as $\gdirac$ and given by the Breit formula \cite{breit:1928:649}:
%

%
% \subsection{Furry picture}
% Within the Furry picture, 
The one-electron wave function obeys the Dirac equation,
\be
\label{eq:Dirac}
  h^{\rm D} \ket{a} = \varepsilon_a \ket{a}
\,,\qquad
  h^{\rm D} = -i \balpha \cdot \bnabla + \beta + V(r)
% \balpha \cdot \bp + \beta + V_{\rm {eff}} (\br) 
\,,
\ee
where the binding potential $V(r)$ includes the nuclear potential and optionally some effective screening potential. 
Within the independent-electron approximation, the energy shift $\Delta E$ is found as an expectation value of $V_{\rm m}$ with $\ket{a}$, which yields the following expression for the $g$ factor,
% For systems with one electron over closed shells in the one-electron approximation the value of the $g$-factor value can be found as
%
\be
\label{eq:gbreit}
 g^{(0)} = \dfrac{1}{m_a}\langle a \vert U \vert a \rangle 
\,.
\ee
%
%where $\vert a\rangle$ denotes a one-electron state.
% 
For the pure Coulomb nuclear potential, we denote the $g$ factor as $g^{(0)}_\mathrm{C}$. In the case of the point nucleus, $g^{(0)}_\mathrm{C}$ is known analytically (we denote it as $\gdirac$) and for the $2s$ state given by the Breit formula \cite{breit:1928:649}:
% %
 \be
 \label{eq:gbreit_point}
   \gdirac = \dfrac{2}{3}(1 + \sqrt{2+2\gamma}) = 2 - \dfrac{(\alpha Z)^{2}}{6} + \dots
 \,,
 \ee
% %
where $\gamma = \sqrt{1 - (\alpha Z)^{2}}$.
%We denote by $g^{(0)}_\mathrm{C}$ in Eq.~\eqref{eq:gbreit_point} the value obtained for the Coulomb nuclear potential.

The total $g$-factor value comprises $\gdirac$ and various corrections,
\be
\label{eq:gtotal}
  g = \gdirac + \dgint + \dgqed + \dgnuc
\,.
\ee
Here, $\dgint$ is the interelectronic-interaction correction which is the main topic of this work, $\dgqed$ is the QED correction previously investigated for lithiumlike ions, e.g., in Refs.~\cite{yerokhin:2002:143001, yerokhin:2004:052503, glazov:2006:330, volotka:2009:033005, glazov:2010:062112, andreev:2012:022510, volotka:2014:253004, yerokhin:2017:060501, cakir:2020:062513, yerokhin:2020:022815, kosheleva:2022:103001}, $\dgnuc$ stands for the nuclear size \cite{karshenboim:2000:380, glazov:2002:408, yerokhin:2013:245002}, nuclear recoil~\cite{shabaev:2002:091801, koehler:2016:10246, shabaev:2017:263001, malyshev:2017:731, shabaev:2018:032512} and nuclear polarization~\cite{nefiodov:2002:081802, volotka:2014:023002} effects.  

%was calculated rigorously (to all orders in $\alpha Z$) up to the second order in $1/Z$~\cite{volotka:2012:073001, wagner:2013:033003, volotka:2014:253004, yerokhin:2021:022814}. The QED correction also includes contributions rigorously calculated, namely one-loop, one-electron and two-electron contributions. The higher-order QED contributions are known only within some approximation so far~\cite{pachucki:2004:150401, pachucki:2005:022108, jentschura:2009:044501, yerokhin:2013:042502, czarnecki:2016:060501, czarnecki:2018:043203, czarnecki:2020:050801,aoyama:2019:28,shabaev:2002:062104,shabaev:2002:062104,glazov:2004:062104}. 

In the present work we focus on the electronic-structure contribution $\Delta g_{\rm int}$ to the ground-state $g$ factor of lithiumlike ions. 
%
%This correction is accounted for by QED perturbation theory, with the interaction Hamiltonian $H_I$ equal to a sum of the $H_{\rm QED}$ and $V^{\rm magn}$. The term $H_{\rm QED}$ is the usual QED Hamiltonian \cite{mohr:1998:227} describing the interaction between the electron-positron field and the photon field and $V^{\rm magn}$ is the operator describing the interaction of bound electrons with external magnetic field, see Eq.~\eqref{eq: V_magn}.
%
The evaluation procedure for $\Delta g_{\rm int}$ is described in the following section. Here, we discuss an important aspect of this procedure --- the choice of the zeroth-order approximation, which is defined by the potential $V(r)$ in the Dirac equation (\ref{eq:Dirac}).
In the original Furry picture it is the Coulomb potential $V_{\rm C}(r)$ generated by the nucleus. So, the interelectronic interaction is completely neglected at this stage.
In present work we consider the extended Furry picture, which is based on the Dirac equation in the presence of an effective potential $V_{\rm eff}(r)$,
\begin{equation}
  V_{\rm eff}(r) = V_{\rm C}(r) + V_{\rm scr}(r)
\,,
\label{eq:V_eff} 
\end{equation}
where $V_{\rm scr}(r)$ is some local screening potential. This approach accelerates the convergence of perturbation theory, at least within the several leading orders, by accounting for a part of the interelectronic interaction already in the zeroth order. Another advantage of the screening potential is that it lifts the near-degeneracy of the $(1s)^2 2s$ and $(1s)^2 2p_{1/2}$ states occurring for the Coulomb potential. This helps to avoid the problems with additional singularities arising from the corresponding intermediate states, which are especially nuisance in the QED calculations involving energy integration. In particular, a subtraction procedure was used in Ref.~\cite{yerokhin:2021:022814} in order to deal with these singularities for the Coulomb potential.
%Moreover, in this way we relieve the quasidegeneracy of the $1s^2 2s$ and $1s^2 2p_{1/2}$ states already at the zeroth-order level and improve the energy level scheme of the first excited states. 
%

Following our previous investigations, we employ 5 binding potentials: Coulomb, core-Hartree, Dirac-Hartree, Kohn-Sham, and Dirac-Slater. 
A well-known choice of $V_{\rm eff}(\br)$ is the core-Hartree (CH) potential
\be
\label{eq:CH}
V_{\rm eff}(r) = V_{\rm C}(r) + \alpha \int^{\infty}_{0} dr' \dfrac{1}{r_{>}} \rho_c (r').
\ee
Here $\rho_c$ is the density of the core (closed shells) electrons,
\be
\label{eq:dens}
\rho_c (r') = \sum_{\kappa_c, n_c} (2j_c + 1)(G^2_c(r) + F^2_c(r)),
\ee
where $\kappa_c$ and $n_c$ are the quantum numbers of the closed shells, $G_c$ and $F_c$ are the corresponding radial components of the wave function. The potential derived from the density functional theory reads
\be
\label{eq:dft}
V_{\rm eff}(r) = V_{\rm C}(r) + \alpha \int^{\infty}_{0} dr' \dfrac{1}{r_{>}} \rho_t (r') - x_{\alpha} \dfrac{\alpha}{r} \left(\dfrac{81}{32\pi^2}r \rho_t(r) \right)^{1/3}.
\ee
Here $\rho_t$ is the total electron density, including the closed shells and the valence electron,
\be
\label{eq:denstot}
\rho_t (r) = (G^2_a(r) + F^2_a(r)) + \sum_{\kappa_c, n_c} (2j_c + 1)(G^2_c(r) + F^2_c(r)).
\ee
The parameter $x_\alpha$ varies from $0$ to $1$. The cases of values $x_\alpha = 0, \, 2/3$,
and $1$ are referred to as the Dirac-Hartree (DH), Kohn-Sham (KS), and Dirac-Slater (DS) potentials, respectively.

The DFT potentials as given by Eq.~(\ref{eq:dft}) possess non-physical asymptotic behavior. The Latter correction~\cite{Latter:1955:510} circumvents this problem, but as a consequence the potentials cease to be smooth. The smoothing procedure itself can be different and therefore the potentials are barely reproducible. In the present work, we compare the total results for the potentials with and without the Latter correction and demonstrate that the corresponding shifts are well within the scatter between potentials. This is what one would expect for the sum of the perturbation-theory expansion. At the same time, the values without the Latter correction have better convergence with respect to the number of basis functions due to the potential smoothness. For these reasons, we propose to use this option once the higher-order perturbation theory terms are taken into account. 
%
%\begin{itemize}
%\item We note, when constructing the screening potentials it is important to take care about the asymptotic behaviour of a potential $(Z-2)/r$
    
%In present work we use potentials without Latter correction because this correction is defined differently by authors (hard to compare), numerical integration is bad. It can be explained by the fact that these potentials are smooth functions of r. Later we show, that the total result is the same. It was found that results obtained within modified potentials better converge over the number of basis functions. 
  
%\end{itemize}
%
%{\rd In order to avoid a double-counting of the screening effects, the counterterm $-V_{\rm scr}(\br)$ should be added to the interaction Hamiltonian, thereby the additional counterterm diagrams appear.}
% 
%Alternatively, one can introduce an effective screening potential in the Dirac equation (extended Furry picture), which leads to the corresponding ``screened'' value of $g^{(0)}$.
\indent

%
% ======================== MANY-ELECTRON EFFECTS =======================
%
\section{Many-electron effects}
\label{sec:me}
In the framework of bound-state QED perturbation theory the interelectronic-interaction contribution $\Delta g_{\rm int}$ can be written as:
\be
\label{eq:g_int}
  \Delta g_{\rm int} &=& \Delta g^{(0)}_{\rm int} + \Delta g^{(1)}_{\rm int} + \Delta g^{(2)}_{\rm int} + \Delta g^{(3+)}_{\rm int}
\,,
\ee
where $\Delta g^{(i)}_{\rm int}$ is the $i$th order correction in $\alpha$, namely $\Delta g^{(1)}_{\rm int}$ and $\Delta g^{(2)}_{\rm int}$ are the corrections to the bound-electron $g$ factor due to the one- and two-photon exchange, respectively, $\Delta g^{(3+)}_{\rm int}$ denotes the sum of the higher-order corrections. The zeroth-order term $\Delta g^{(0)}_{\rm int}$ is just the difference between the one-electron values in the extended and original Furry picture,
\be
\label{eq:g_scr}
  \Delta g^{(0)}_{\rm int} &=& g^{(0)} - g^{(0)}_\mathrm{C}
\,.
\ee
Below, we consider the evaluation of the first- and second-order terms within the rigorous QED approach and of the higher-order part within the Breit approximation. 
% 
% In the framework of the extended Furry picture, one can obtain zeroth-order value  $g^{(0)}$ using Eq.~\eqref{eq:gbreit}. %with wave function  $\vert a\rangle$ constructed from the one-electron wave-function being a solution of Dirac equation with some effective potential. 
% We note, that $g^{(0)}$ already partially includes interelectronic-interaction effects, therefore, within extended Furry picture, the additional term $\Delta g^{(0)}_{\rm int}$ in expansion Eq.~\eqref{eq:g_int} appears:
%
%
%-----------------------------------------------------------------------
%
\subsection{First-order contribution}
\label{subsec:A}
\begin{figure}
\begin{center}
\includegraphics[width=0.25\linewidth]{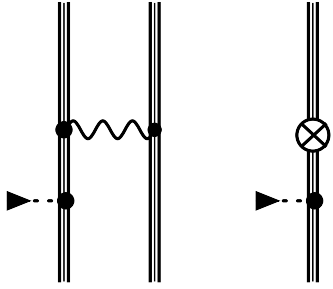}
\caption{Feynman diagrams representing the one-photon-exchange correction to the $g$ factor in the framework of the extended Furry picture. The wavy line indicates the photon propagator. The triple line represents the electron propagator in the effective potential $V_{\rm eff}$. The dashed line terminated with the triangle corresponds to the interaction with external magnetic field. The symbol $\otimes$ represents the screening potential counterterm.}
\label{ris:one_ph_ex}
\end{center}
\end{figure}
%
%\\
% \indent
%
Within the framework of bound-state QED, each term of the perturbation theory is represented by the corresponding set of diagrams. Currently, there are several methods to derive formal expressions from the first principles of QED: the two-time Green's function method \cite{shabaev:2002:119}, the covariant-evolution-operator method \cite{lindgren:2004:161}, and the line profile approach \cite{andreev:2008:135}. The corresponding formulas for $\Delta g^{(i)}_{\rm int}$ were derived in Ref.~\cite{shabaeva:1995:2811} (first order) and in Ref.~\cite{volotka:2012:073001} (second order) within the two-time Green's function method.
The leading correction $\Delta g^{(1)}_{\rm int}$ to the $g$ factor corresponds to the one-photon-exchange diagram in the presence of magnetic field, see Fig.~\ref{ris:one_ph_ex}. Also the counterterm diagram appears in the presence of the screening potential. This diagram is also shown in Fig.~\ref{ris:one_ph_ex}, where the symbol $\otimes$ denotes the counterterm.
The corresponding expression for the interelectronic-interaction correction reads as:
\begin{align}
\label{eq:1ph}
  \Delta g^{(1)}_{\mathrm{int}} = \dfrac{1}{m_a} \sum_{b} \sum_{P,Q}(-1)^{P+Q}\Bigg[ 2 {\sum_{n}}'
    \frac{\matrixel{Pa\,Pb}{I(\veps_{Qa}-\veps_{Pa})}{n\,Qb}\matrixel{n}{U}{Qa}}
         {\veps_{Qa}-\veps_{n}} 
%\nonumber\\
%+         
%         2\sum_{P}(-1)^{P}{\sum_{n}}'
%         \frac{\matrixel{PaPb}{I(\Delta_{Paa)}}{an}\matrixel{n}{U}{b}}
%         {\veps_{b}-\veps_{n}} 
\nonumber\\
    - \matrixel{Pa\,Pb}{I'(\veps_{Qa}-\veps_{Pa})}{Qa\,Qb}\matrixel{Qb}{U}{Qb} \Bigg]
    - \dfrac{1}{m_a} {\sum_{n}}' \frac{\matrixel{a}{V_{\rm scr}}{n}\matrixel{n}{U}{a}} {\veps_{a}-\veps_{n}}
\,.
\end{align}
%
%\be
%\label{B(aZ)}
% \Delta g^{(1)} &=& \frac{2}{\Delta E_a} \, \sum_b  \Bigg[  \sum_P (-1)^P \Biggl( \la \zeta_{b\vert PaPb} | V^{\rm mag} | a \ra + \la \zeta_{a\vert PbPa} | V^{\rm mag} | b \ra \Bigg) 
% \nonumber\\ 
% &-& \dfrac{1}{2}\Bigg( \la a | V^{\rm mag} | a \ra - \la b | V^{\rm mag} | b \ra \Bigg) \la a b | I^{\prime}(\veps_a-\veps_b) | b a \ra \Bigg] - \frac{2}{\Delta E_a}\,\la \eta_a  | V^{\rm mag} | a \ra
%\ee
%
Here $P$ and $Q$ are permutation operators giving rise to the sign $(-1)^{P+Q}$ according to the parity of the permutation, $\veps_n$ are the one-electron energies, $\vert b \rangle$ stands for the $1s$ state, while the summation over $b$ runs over two possible projections $m_b = \pm 1/2$. The prime on the sums over the intermediate states $n$ denotes that the terms with vanishing denominators are omitted. 

The interelectronic-interaction operator $I(\omega)$ in the Feynman gauge is given by
\begin{equation}
\label{eq:ii_feyn}  
  I(\omega, r_{12} ) = \alpha (1-\balpha_{1} \cdot \balpha_{2}) \dfrac{\exp{(i\tomega r_{12})}}{r_{12}}
\,.
\end{equation}
In the Coulomb gauge it is given by
\begin{equation}
\label{eq:ii_coul}  
  I(\omega, r_{12}) = \alpha \left( \dfrac{1}{r_{12}} 
    - \balpha_{1} \cdot \balpha_{2} \dfrac{\exp{(i\tomega r_{12})}}{r_{12}}
    - \left[\balpha_{1} \cdot \bnabla_{1},\left[\balpha_{2} \cdot \bnabla_{2},
      \dfrac{\exp{(i\tomega r_{12})}-1}{\omega^{2}r_{12}}\right] \right] \right)
\,.
\end{equation}
Here $r_{12}= \vert \bfr_1-\bfr_2 \vert$, and $\tomega=\sqrt{\omega^2+i0}$, the branch of the square root is fixed by the condition ${\rm Im}\,\tomega>0$. In Eq.~(\ref{eq:1ph}) the notation $I^{\prime}(\omega) = dI(\omega)/d\omega$ is used. 

Assuming $\omega = 0$ in the Coulomb gauge we obtain $I$ in the Breit approximation,
\begin{equation}
  I_{\rm B}(r_{12}) = \alpha \left( \dfrac{1}{r_{12}} 
    - \frac{\balpha_{1} \cdot \balpha_{2}}{r_{12}} 
    + \frac{1}{2}\left[\balpha_{1} \cdot \bnabla_{1},\left[\balpha_{2} \cdot \bnabla_{2},r_{12}\right] \right]
  \right)\,.
\label{eq:breit0} 
\end{equation}
This form is used to calculate the third- and higher-order contributions, see Sec.~\ref{subsec:C}.
% the one-photon-exchange correction to the $g$ factor in Li-like ions
% Note that rigorous evaluation of the one-photon exchange correction to the $g$ factor in Li-like ions was previously performed in Ref.\cite{shabaev:2002:062104}, for example.
%
%-----------------------------------------------------------------------
%
\subsection{Second-order contribution}
\label{subsec:B}

\begin{figure}
\begin{center}
\includegraphics[width= 0.7\linewidth]{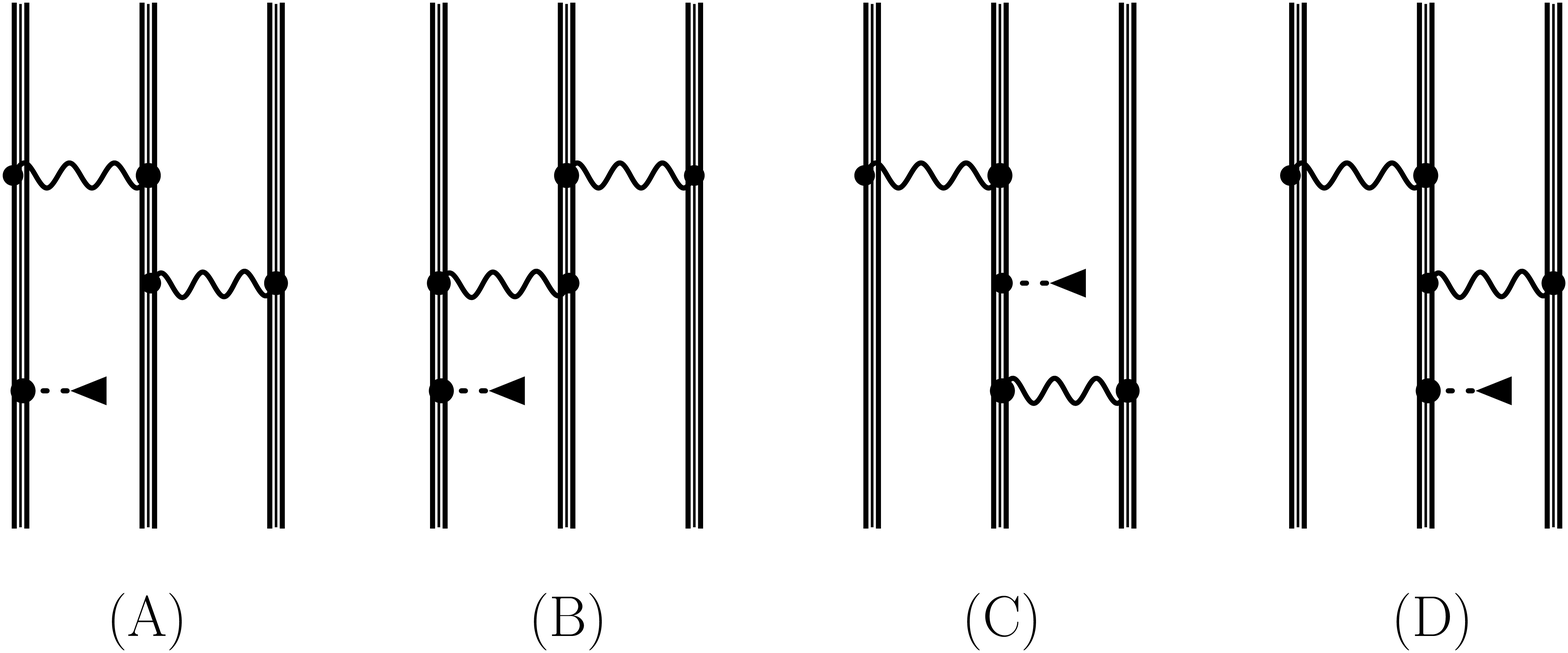}
\caption{Feynman diagrams representing the three-electron part of the two-photon-exchange correction to the $g$ factor in the framework of the extended Furry picture. Notations are the same as in Fig.~\ref{ris:one_ph_ex}.}
\label{fig-3el}
\end{center}
\end{figure}

\begin{figure}
\begin{center}
\includegraphics[width= 0.7\linewidth]{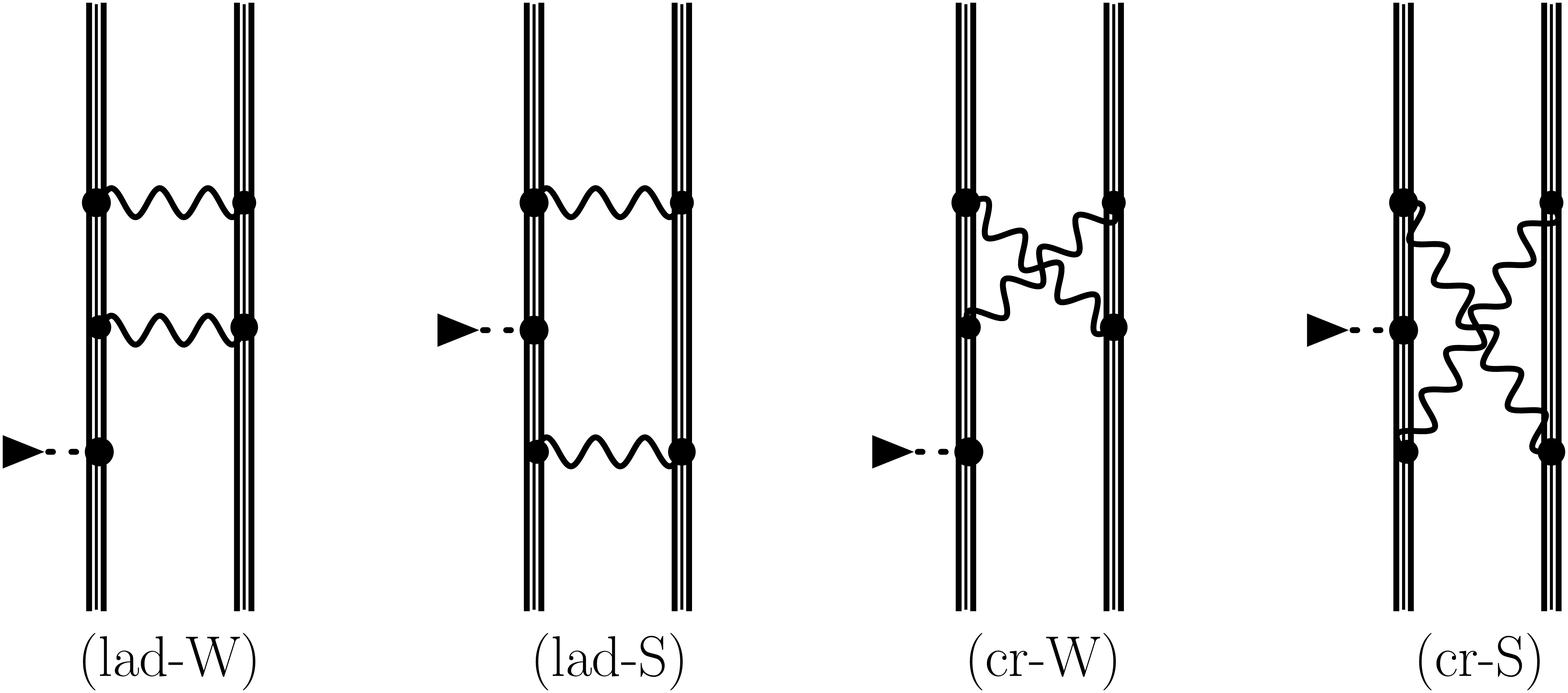}
\caption{Feynman diagrams representing the two-electron part of the two-photon-exchange correction to the $g$ factor in the framework of the extended Furry picture. Notations are the same as in Fig.~\ref{ris:one_ph_ex}.}
\label{fig-2el}
\end{center}
\end{figure}

\begin{figure}
\begin{center}
\includegraphics[width= 0.8\linewidth]{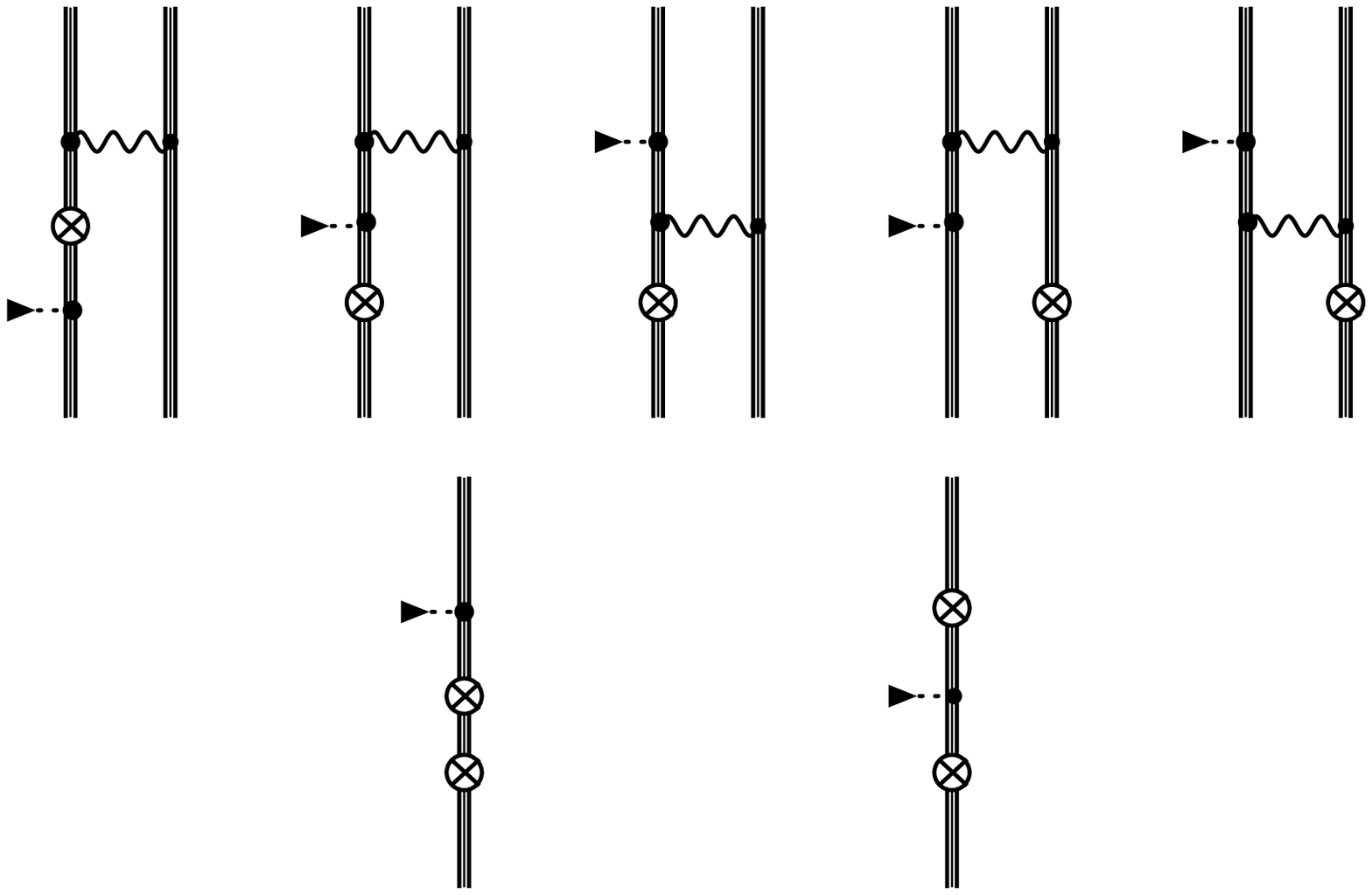}
\caption{Feynman diagrams representing the counterterm part of the two-photon-exchange correction to the $g$ factor in the framework of the extended Furry picture. Notations are the same as in  Fig.~\ref{ris:one_ph_ex}.}
\label{fig-countertems}
\end{center}
\end{figure}

The second-order correction $\Delta g^{(2)}_{\rm int}$ corresponds to the two-photon-exchange diagrams which can be divided into two large classes, namely three-electron (Fig. \ref{fig-3el}) and two-electron (Fig. \ref{fig-2el}) ones. In the extended Furry picture, the counterterm diagrams shown in Fig. \ref{fig-countertems} appear in addition. The contributions of these diagrams are divided into reducible and irreducible parts. Reducible part is the contributions in which the energies of the intermediate and reference states coincide, while the irreducible part is the remainder. The reducible part is taken together with the non-diagrammatic perturbation-theory terms of the corresponding order.

So, the second-order correction $\Delta g^{(2)}_{\rm int}$ to the $g$ factor can be written as follows,
\begin{equation}
\label{eq:2ph}
  \Delta g^{(2)}_{\rm int} = \Delta g^{(2)}_{\rm 3el} + \Delta g^{(2)}_{\rm 2el} + \Delta g^{(2)}_{\rm ct} + \Delta g^{(2)}_{\rm red}
\,.
\end{equation}
%
% with three-electron $\Delta g^{(2)}_{\rm 3el}$, two-electron $\Delta g^{(2)}_{\rm 2el}$, counterterm $\Delta g^{(2)}_{\rm ct}$, and reducible $\Delta g^{(2)}_{\rm red}$ contributions.
%
%
% \\ 
% \indent
%
The three-electron contribution $\Delta g^{(2)}_{\rm 3el}$ can be written as the sum,
\begin{equation}
\label{eq:2ph_3el}
  \Delta g^{(2)}_{\rm 3el} = \Delta g^{(2)}_{\rm 3el, A} + \Delta g^{(2)}_{\rm 3el, B} + \Delta g^{(2)}_{\rm 3el, C} + \Delta g^{(2)}_{\rm 3el, D}\,,
\end{equation}
where each term is the irreducible part of the corresponding Feynman diagram in Fig. \ref{fig-3el}.
% : the diagrams (A) and (B) with one or two vertices on one electron line, while contributions (C) and (D) denote diagrams with three vertices on one electron line (Appendix A).
The formal expressions for $\Delta g^{(2)}_{\rm 3el}$ can be found in Appendix \ref{sec:A:A}, they involve double summation over the Dirac spectrum.
The two-electron contribution $\Delta g^{(2)}_{\rm 2el}$, in contrast to the $\Delta g^{(2)}_{\rm 3el}$, comprises triple summation and the integration over the virtual photon energy $\omega$, thus making its evaluation significantly more involved, including development of the numerical procedure. Similarly to Eq.~\eqref{eq:2ph_3el}, we represent it in the following form,
\begin{equation}
\label{eq:2ph_2el}
  \Delta g^{(2)}_{\rm 2el} = \Delta g^{(2)}_{\rm 2el, lad-W} + \Delta g^{(2)}_{\rm 2el, lad-S} + \Delta g^{(2)}_{\rm 2el, cr-W} + \Delta g^{(2)}_{\rm 2el, cr-S}
\,.
\end{equation}
%
%\ss
This contribution consist of ladder (``lad'') and cross (``cr'') parts, see Fig.~\ref{fig-2el}, which are named by analogy with the two-photon-exchange diagrams without the external-field vertex \cite{shabaev:2002:119}, the labels ``W'' and ``S'' indicate the position of this vertex. The formal expressions for these terms can be found in Appendix \ref{sec:A:B}.

The third term $\Delta g^{(2)}_{\rm red}$ in Eq.~\eqref{eq:2ph} includes the reducible parts of all diagrams, both two-electron and three-electron, including the non-diagrammatic terms, see Appendix \ref{sec:A:C}.
Finally, the counterterm contribution $\Delta g^{(2)}_{\rm ct}$ corresponds to the diagrams in Fig.~\ref{fig-countertems} which arise when the screening potential is included in the Dirac equation. 
%
%The formal expression of each term of Eq.~\eqref{eq:2ph} can be found in Appendix.
%To avoid  double-counting  of  the screening  effects,  one  has  to  add  the  counterterm  to  the interaction  Hamiltonian  and  to  evaluate  the  contribution  of the corresponding diagrams.

%arises from the formal expressions for the mentioned above corrections

%
% The second-order correction $\Delta g^{(2)}$ to the $g$ factor is represented by the two-photon exchange diagrams Figs.~\ref{fig-2el}, \ref{fig-3el}, and \ref{fig-countertems}.
%
%These diagrams are divided into three groups: the two-electron (Fig.~\ref{fig-2el}), the three-electron (Fig.~\ref{fig-3el}), and the counterterm (Fig.~\ref{fig-countertems}) ones.
%\begin{equation}
%\Delta g^{(2)} = \Delta g^{(2)}_{\rm 3el} + \Delta g^{(2)}_{\rm 2el} + \Delta g^{(2)}_{\rm countrterms} + \Delta g^{(2)}_{\rm reducible}\,,
%\end{equation}
%
%The formal expressions for the first two groups were firstly derived in Ref.~\cite{volotka:2012:073001} for the case of original Furry picture. As was already mentioned above, in the extended Furry picture additional counterterm diagrams have to be taken into account. The formal expressions corresponding to the counterterm diagrams depicted in Fig.~\ref{fig-countertems} as well as two- and three-electron diagrams can be found in Appendix.
%
\begin{table*}
\caption{\label{T1_gauges} Individual contributions to the two-photon-exchange correction for the ground-state $g$ factor of $^{28}$Si$^{11+}$ and $^{208}$Pb$^{79+}$ obtained for the core-Hartree potential in the Feynman and Coulomb gauges.}
\centering
\setlength{\tabcolsep}{5pt}
\resizebox{1.0\columnwidth}{!}{
\begin{tabular}{l S S S S} 
\hline
\hline
&\multicolumn{2}{c}{$^{28}$Si$^{11+}$}	
&\multicolumn{2}{c}{$^{208}$Pb$^{79+}$} 	 \\
Contribution       
&\multicolumn{1}{c}{Feynman}	
&\multicolumn{1}{c}{Coulomb} 
&\multicolumn{1}{c}{Feynman}	
&\multicolumn{1}{c}{Coulomb}	 \\ 
\hline
3el, A             &   -35.5174    &   -32.1510    &  -47.5059    &  -57.4179    \\
3el, B             &    39.5459    &    36.1787    &   50.7785    &   60.6392    \\
3el, C             & 19540.5865    & 19537.1073    &  625.5797    &  621.2035    \\
3el, D             &   -96.6335    &   -96.2590    & -120.4084    & -122.3388    \\
2el, lad-W         &-13638.4718    &-13663.2943    &-3016.6610    &-3054.2046    \\
2el, cr-W          &   -83.2577    &   -60.5589    &-2679.1832    &-2648.0997    \\
2el, lad-S         & 20478.7090    & 20571.8104    &  606.4759    &  614.7309    \\
2el, cr-S          &    22.9387    &    -0.4702    &   12.8076    &   -1.0717    \\
red, E             &-19546.8105    &-19543.7536    & -627.3047    & -622.8958    \\
red, F             &   234.1531    &   234.2034    &  296.3070    &  298.5116    \\
red, G             &     0.9138    &    -0.0331    &    0.7037    &   -1.6609    \\
red, 2el           & -6923.7189    & -6990.3424    & 4889.2411    & 4903.4344    \\
%reducible         &-26235.4625    &-26299.9257    && \\
ct-1               &    18.6909    &    18.6909    &   20.0834    &   20.0834    \\ 
ct-2               &   -10.9920    &   -10.9920    &  -11.0497    &  -11.0497    \\ 
%counterterms      &     7.6989    &     7.6989    && \\ 
  Total            &     0.1362(1) &     0.1362(1) &   -0.1361(1) &   -0.1361(1) \\ \hline\hline 
\end{tabular}}
\end{table*}
%
%-----------------------------------------------------------------------
%
\subsection{Higher-order contribution}
\label{subsec:C}
Rigorous evaluation of the higher-order term $\Delta g^{(3+)}_{\rm int}$ to all orders in $\alpha Z$ is not feasible at the moment.
In this case, one of the currently available methods \cite{bratsev:1977:2655, dzuba:1987:1399, blundell:1989:2233, shabaev:1995:3686, boucard:2000:59, zherebtsov:2000:701, yerokhin:2008:R020501, ginges:2018:032504} can be considered. 
% Once the two-photon-exchange diagrams were evaluated within the rigorous QED approach~\cite{volotka:2014:253004}, the
The all-order CI-DFS method \cite{bratsev:1977:2655} was employed in Refs.~\cite{glazov:2004:062104,glazov:2005:55} to find the contribution of the second and higher orders. In Ref.~\cite{volotka:2014:253004} the two-photon-exchange diagrams were evaluated within the rigorous QED approach, and thus only $\Delta g^{(3+)}_{\rm int}$ was needed from the CI-DFS calculations. This was a demanding task since the subtraction of the leading orders from the total value delivered by any all-order method requires high enough numerical accuracy. In this paper, we use the recursive formulation of perturbation theory~\cite{glazov:2017:46}, which provides an efficient way to accesses the individual terms of the perturbation expansion up to any order, in principle. Application of this method to the $g$ factor of lithiumlike silicon and calcium in Refs.~\cite{glazov:2019:173001, kosheleva:2022:103001} has demonstrated the ability to provide significantly better accuracy than the CI-DFS method. Recently, non-relativistic quantum electrodynamics (NRQED) approach was used to solve this task with even better accuracy~\cite{yerokhin:2020:022815, yerokhin:2021:022814}. At the same time, perturbation theory allows one to include screening potential which is an important ingredient here~\cite{kosheleva:2022:103001}, while the NRQED calculations were based on the Coulomb potential so far.
%
%The interelectronic interaction within the Breit approximation can be taken into account within any of the available methods \cite{boucard:2000:59, dzuba:1987:1399, ginges:2018:032504, shabaev:1995:3686, zherebtsov:2000:701, yerokhin:2008:R020501, blundell:1989:2233, bratsev:1977:2655}. Previously, the all-order CI-DFS method \cite{bratsev:1977:2655} was employed to evaluate $\Delta g_{\rm int}^{(3+)}$ \cite{volotka:2012:073001}. In this work, we opt for the recursive formulation of the perturbation theory \cite{glazov:2017:46}. This method allows one to access efficiently the individual terms of the perturbation expansion up to any order. It also easily ensures that the zeroth-order Hamiltonian is the same for the rigorous QED and Breit-approximation calculations. Recently, this method has been successfully applied to the similar calculations of the higher-order contributions to the $g$ factor of Li-like ions \cite{glazov:2019:173001}.

To formulate our approach, we start with the Dirac-Coulomb-Breit equation,
\begin{equation}
\label{eq:DCB}
  \Lambda_+ \left( H_0 + H_1 \right) \Lambda_+ \ket{A} = E_A \ket{A}
\,,
\end{equation}
where $\Lambda_+$ is the projection operator constructed as the product of one-electron projectors on the positive energy states. Zeroth-order Hamiltonian $H_0$ is the sum of the one-electron Dirac Hamiltonians,
\begin{equation}
\label{eq:H_0}
  H_0 = \sum_j h^{\rm D}(j)
\,,
\end{equation}
where $h^{\rm D}$ is given by Eq.~\eqref{eq:Dirac}.
% $h^{\rm D} = \balpha \cdot \bp + \beta + V_{\rm {eff}} (\br)$. 

Let us introduce the zeroth-order eigenfunctions $N^{(0)}$ as 
\begin{equation}
\label{eq:N_0}
  \Lambda_+ H_0 \Lambda_+ \ket{N^{(0)}} = E_N^{(0)} \ket{N^{(0)}}
\,.
\end{equation}
These functions form an orthogonal basis set of many-electron wave functions. In our case, they are constructed as the Slater determinants of one-electron solutions of the Dirac equation. In particular, for the reference state $\ket{A}$ in the zeroth approximation we have
\begin{equation}
\label{eq:A_0}
  \Lambda_+ H_0 \Lambda_+ \ket{A^{(0)}} = E_A^{(0)} \ket{A^{(0)}}
\,.
\end{equation}
The perturbation $H_1$ in \eqref{eq:DCB} reads as
\begin{equation}
  H_1 = \,\sum_{j<k} I_{\rm B}(j,k) - \sum_j V_{\rm scr}(j)
\,,
\end{equation}
and represents the interelectronic interaction in the Breit approximation with the screening potential subtracted.

We use the perturbation theory with respect to $H_1$, which yields the following expansions for the energy $E_A$ and the wave function $\ket{A}$,
\begin{align}
\label{eq:ten}
  E_A &= \sum_{k=0}^{\infty} E_A^{(k)}
\,,\\
\label{eq:ta}
  \ket{A}
   &= \sum_{k=0}^{\infty} \ket{A^{(k)}}
    = \sum_{k=0}^{\infty} \sum_N \ket{N^{(0)}} \braket{N^{(0)}}{A^{(k)}}
\,.
\end{align}
In Ref.~\cite{glazov:2017:46} the recursive scheme to evaluate $E_A^{(k)}$ and $\braket{N^{(0)}}{A^{(k)}}$ order by order was presented. We emphasize that instead of the widely used normalization $\braket{A^{{(0)}}}{A}=1$, we impose the condition $\braket{A}{A}=1$. Below we consider how to use this method to find the interelectronic-interaction contributions $\Delta g_{\rm int}^{(k)}$.

For the many-electron state $\vert A\rangle$ the $g$ factor is found as 
\be
\label{eq:g_A}
  g = \dfrac{1}{m_a} \langle A \vert \sum_i U(i) \vert A \rangle 
\,,
\ee
and in the zeroth approximation with $\vert A^{(0)}\rangle$, this expression reduces to Eq.~\eqref{eq:gbreit}.
% Accordingly, $\Delta g_{\rm int}$ is given by 
% %
% \begin{equation}
% \label{eq:XA}
%   \Delta g_{\rm int} = \dfrac{1}{m_a} \, \matrixel{A}{\sum_i U(i)}{A} - g^{(0)}_\mathrm{C}\,,
% \end{equation}
% %
% from where, substituting Eq.~(\ref{eq:ta}) we find the $k$-th order contribution as
Substituting the expansion \eqref{eq:ta} we find the term of the order $k\geq 1$ as
\begin{align}
\label{eq:XAk}
  \Delta g_{\rm int}^{(k)}[+] &= \dfrac{1}{m_a} \, \sum_{j=0}^{k} \matrixel{A^{(j)}}{\sum_i U(i)}{A^{(k-j)}}
\nonumber\\
  &= \dfrac{1}{m_a} \, \sum_{j=0}^{k} \sum_{M,N} \braket{A^{(j)}}{M^{(0)}} \matrixel{M^{(0)}}{\sum_i U(i)}{N^{(0)}} \braket{N^{(0)}}{A^{(k-j)}}
\,.
\end{align}
In fact, Eq.~\eqref{eq:XAk} yields the contribution of the positive-energy states only, since only positive-energy excitation are included in the expansion \eqref{eq:ta}. For this reason, we label the result with ``$[+]$'' sign. However, the negative-energy spectrum is equally important, see e.g. Refs.~\cite{glazov:2004:062104, maison:2019:042506}. Its contribution is given by the following expression,
\begin{equation}
\label{eq:XAneg}
  \Delta g_{\rm int}[-] = \dfrac{2}{m_a} \, \sum_{p,n} \frac{\matrixel{p}{U}{n}}{\veps_p-\veps_n}
             \matrixel{\hat a^+_n \hat a_p A}{H_1}{A}
\,.
\end{equation}
Here $\ket{p}$ and $\ket{n}$ are the positive- and negative-energy one-electron states, respectively, $\hat a^+$ and $\hat a$ are the corresponding creation and annihilation operators. The term of the order $k$ reads,
\begin{equation}
\label{eq:XAnegk}
  \Delta g_{\rm int}^{(k)}[-] = \dfrac{2}{m_a} \, \sum_{j=0}^{k-1} \sum_{M,N} \braket{A^{(j)}}{M^{(0)}}
    \left[ \sum_{p,n} \frac{\matrixel{p}{U}{n} \matrixel{\hat a^+_n \hat a_p M^{(0)}}{H_1}{N^{(0)}}}{\veps_p-\veps_n} \right] 
  \braket{N^{(0)}}{A^{(k-j-1)}}
\,.
\end{equation}
%
% from the coefficients of the wave function $\braket{N^{(0)}}{A^{(k)}}$ obtained within the recursive scheme
The equations~\eqref{eq:XAk} and \eqref{eq:XAnegk} provide the interelectronic-interaction contributions within the Breit approximation,
\begin{equation}
\label{eq:XAnegka}
  \Delta g_{\rm int, \rm L}^{(k)}= \Delta g_{\rm int}^{(k)}[+] + \Delta g_{\rm int}^{(k)}[-] 
\,.
\end{equation}
The complete $k$-order terms can be presented in the following form, 
% We present the calculated contributions in the following form:
%
\be
  \Delta g^{(k)}_{\rm int} = \Delta g^{(k)}_{\rm int, \rm L} + \Delta g^{(k)}_{\rm int, \rm H}
\,,
\ee
%
% where $\Delta g^{(k)}_{\rm int, \rm L}$ denotes the leading-order part~\eqref{eq:XAnegka} and 
where $\Delta g^{(k)}_{\rm int, \rm H}$ is the presently unknown higher-order part, which has to be estimated somehow to ascribe the uncertainty to $\Delta g^{(k)}_{\rm int}$. 
In the present work, $\Delta g^{(3+)}_{\rm int, \rm H}$ is estimated as $\pm 2\Delta g^{(2)}_{\rm int, \rm H} / Z$, as proposed in our previous work \cite{kosheleva:2022:103001}.
\begin{table*}
\caption{\label{T2_Breit} Interelectronic-interaction contributions $\Delta g^{(k)}_{\rm int, \rm L}$ to the ground-state $g$ factor of $^{28}$Si$^{11+}$, in units of $10^{-6}$. The results are obtained in the Breit approximation with the Coulomb and different screening potentials: core-Hartree (CH), Kohn-Sham (KS), Dirac-Hartree (DH), and Dirac-Slater (DS). The last three are considered with (marked by superscript L) and without the Latter correction.} 
\centering
\setlength{\tabcolsep}{0pt}
\resizebox{1.0\columnwidth}{!}{
\begin{tabular}{c S[group-minimum-digits = 4] S[group-minimum-digits = 4] S[group-minimum-digits = 4] S[group-minimum-digits = 4] S[group-minimum-digits = 4] S[group-minimum-digits = 4] S[group-minimum-digits = 4] S[group-minimum-digits = 4]} 
\hline
\hline
$k$	&\multicolumn{1}{c}{Coul}	&\multicolumn{1}{c}{CH} &\multicolumn{1}{c}{KS}	&\multicolumn{1}{c}{$\mathrm{KS^L}$} & \multicolumn{1}{c}{DH} & \multicolumn{1}{c}{$\mathrm{DH^L}$} & \multicolumn{1}{c}{DS} & \multicolumn{1}{c}{$\mathrm{DS^L}$}	\\ 
\hline
$0$   &               & 348.26678     & 341.44336      & 341.3682       & 417.19816     & 353.1638      & 302.70955      & 329.1102       \\
$1$   & 321.43711     & -33.66323     & -25.28906      & -25.211(1)     &-104.40966     & -39.393(1)    &  14.54644      & -11.879(1)     \\
$2$   &  -6.82589(4)  &   0.14664(3)  &  -1.48024(3)   &  -1.4718(2)    &   2.10245(3)  &   1.1326(3)   &  -2.64579(3)   &  -2.5766(2)    \\
$3$   &   0.1077(41)  &  -0.0469(12)  &   0.0343(22)   &   0.0224(14)   &  -0.1978(22)  &  -0.2192(10)  &   0.1001(23)   &   0.0535(13)   \\
$4$   &  -0.01313(93) &   0.00465(43) &  -0.00142(40)  &  -0.00190(23)  &   0.01618(30) &   0.02695(73) &  -0.00433(19)  &  -0.00368(39)  \\
% $5$   &   0.00112(20) &  -0.00012(14) &   0.000091(53) &   0.000079(96) &  -0.00165(16) &  -0.00316(15) &  -0.000153(45) &  -0.000205(50) \\
$5$   &   0.00112(20) &  -0.00012(14) &   0.00009(5)   &   0.00008(10)  &  -0.00165(16) &  -0.00316(15) &  -0.00015(5)   &  -0.00021(5)   \\
Total & 314.7069(42)  & 314.7078(13)  & 314.7070(22)   & 314.706(2)     & 314.7077(22)  & 314.708(2)    & 314.7058(23)   & 314.704(2)      \\ 
\hline
\hline 
\end{tabular}}
\end{table*}

\section{RESULTS AND DISCUSSIONS}
\label{sec:results}

In this section we discuss the numerical evaluation of all the considered contributions to $\Delta g_{\mathrm{int}}$ and present the results for lithiumlike ions. 
% formulas obtained above
% Let us first discuss in detail the numerical calculations of one- and two-photon exchange. 
All calculations are based on the dual-kinetically-balanced finite-basis-set method \cite{shabaev:2004:130405} for the Dirac equation with the basis functions constructed from B-splines \cite{sapirstein:1996:5213}.
First, the Dirac equation \eqref{eq:Dirac} is solved with one of the considered potentials. Zeroth-order contribution $\Delta g^{(0)}_{\mathrm{int}}$ is then found according to Eq.~\eqref{eq:g_scr}. Evaluation of $\Delta g^{(1)}_{\mathrm{int}}$ was first accomplished in Ref.~\cite{shabaev:2002:062104} and then became a routine procedure \cite{glazov:2004:062104, volotka:2014:253004, yerokhin:2017:062511, glazov:2019:173001, cakir:2020:062513, kosheleva:2022:103001}. In this work, we calculate it according to Eq.~\eqref{eq:1ph} with the chosen screening potentials to a required accuracy, using the comparison between the Feynman and Coulomb gauges as an additional crosscheck. 
% We calculate the first-order correction using Eq.~\eqref{eq:1ph}. 

The second-order correction $\Delta g^{(2)}_{\mathrm{int}}$ is calculated according to Eqs.~\eqref{eq:2ph}, \eqref{eq:2ph_3el}, \eqref{eq:2ph_2el}, and the formulas from Appendix, which involve double and triple summations over the intermediate states. The number of basis functions is increased up to $N = 210$ to achieve clear convergence pattern of the results and then the extrapolation $N\rightarrow\infty$ is performed. The partial wave summation over the relativistic angular quantum number $\kappa=(j+1/2)^{j+l+1/2}$ was terminated at $|\kappa_{\rm max}| = 16$ and the remainder was estimated using least-squares inverse polynomial fitting. Moreover, two-electron contributions involve integration over the virtual photon energy $\omega$, which requires special attention due to the poles and cuts from the electron and photon propagators. We use the integration contour proposed in Ref.~\cite{mohr:2000:052501} based on a Wick rotation. The number of the integration points is varied to achieve the required accuracy.

% For the one-photon exchange a comparison of two different gauges has been made, but as the calculation of the one-photon exchange is well known and straightforward, we do not describe in detail here the numerical calculation of this contribution. Whereas the two-photon exchange is already a nontrivial contribution and therefore we give some numerical results.

For a consistency check, the two-photon exchange correction is calculated within the Feynman and Coulomb gauges, and the difference between the results is found to be well within the numerical uncertainty. The gauge invariance is demonstrated in Table~\ref{T1_gauges}, where the individual terms and the total $\Delta g^{(2)}_{\mathrm{int}}$ values for $^{28}$Si$^{11+}$ and $^{208}$Pb$^{79+}$ obtained with the core-Hartree potential are presented.

Calculations of the third and higher orders $\Delta g^{(3+)}_{\rm int}$ are carried out within the Breit approximation using Eqs.~\eqref{eq:XAk} and \eqref{eq:XAnegk}. On the one hand, these calculations are rather complicated and time-consuming. On the other hand, the higher the order $k$ the smaller $|\Delta g^{(k)}_{\rm int}|$. Therefore these contributions are required with lower relative accuracy and the corresponding $N$ and $|\kappa_{\rm max}|$ are taken much smaller than for the first and second orders. This is an important advantage of the perturbation theory as compared to CI-DFS and other all-order methods. In particular, to achieve the same accuracy as done here within the CI-DFS method, one would need to calculate the matrix element \eqref{eq:g_A} to ten digits.

The convergence of the perturbation theory is illustrated in Table~\ref{T2_Breit}, where the Breit-approximation values of $\Delta g^{(k)}_{\rm int}$ with $k=0\dots 5$ are given for the ground-state $g$ factor of lithiumlike silicon. Calculations are carried out for the Coulomb, core-Hartree and six different DFT potentials: with ($\mathrm{KS^L}$, $\mathrm{DH^L}$, and $\mathrm{DS^L}$) and without (KS, DH, and DS) the Latter correction. Note that the KS, DH, and DS potentials in Ref. \cite{kosheleva:2022:103001} are \textit{with} the Latter correction, so those values would be placed in the ``L'' columns. It can be seen that the results without the Latter correction for the one- and two-photon exchange are at least an order of magnitude more accurate than with it. The Latter correction causes the unsmoothness of potentials, which leads to numerical instability. At the same time, all the total values are in agreement within the uncertainties. So, we choose to opt out of the Latter correction in the following.

In Table~\ref{T3_IIC} we present the interelectronic-interaction contributions $\dgint$ to the ground-state $g$ factor of lithiumlike ions for $Z = 14, 54, 82$. The results are obtained with the Coulomb and different screening potentials: core-Hartree (CH), Kohn-Sham (KS), Dirac-Hartree (DH), and Dirac-Slater (DS). As seen from this Table, the total values obtained at different screening potentials are quite close to each other and overlap within their uncertainties.
%
% It should be noted that for $Z=14$ the values obtained in this work for potentials without Latter correction are in agreement within the uncertainty with the previous values obtained in Ref. \cite{kosheleva:2022:103001}, where the potentials with Latter correction were used.
%
%Averaging over the screening potentials gives the total value of $\dgint$ equal to $314.8100(26),\, 1298.6102(76),\, 2139.3478(116)\,$ for $Z = 14, 54, 82$, respectively.
%
%As one can see, the total value of interlectronic correlation correction obtained within CH potential lays closer to the average final value for all Z, presented in Table III. Therefore in Table IV, the results are presented for the case of CH potential. 
%
%As one can see, all values of interlectronic correlation correction are close to the obtained average result, therefore, for all Z, we decided to choose one potential, namely the core-Hartree (CH) potential.

Table~\ref{T4_res} shows the values of the interelectronic-interaction correction $\Delta g_{\rm int}$ to the $g$ factor of lithiumlike ions for $Z = 14$ -- $82$. We have chosen the CH potential here since it is defined unambiguously in contrast to the DFT potentials. For comparison, the values from Refs.~\cite{volotka:2014:253004,yerokhin:2021:022814} are given. As one can see, we have improved the accuracy by an order of magnitude as compared to Volotka \textit{et al.} (2014). The disagreement with the Coulomb-potential results of Yerokhin \textit{et al.} (2021) awaits further investigation.
The uncertainty of the total values is mainly determined by the numerical error of $\Delta g^{(3+)}_\text{int,L}$ and by estimation of unknown $\Delta g^{(3+)}_\text{int,H}$. The former is obtained by analyzing the dependence of the results on the basis size. The latter is found as $\pm 2\Delta g^{(2)}_{\rm int, \rm H} / Z$.
% was estimated as $1.5 (\alpha Z)^4/Z^3$, where the coefficient $1.5$ was obtained from the fitting of the uncertainties of the total value $\dgint$ for $Z = 14, 54, 82$ by function $a(\alpha Z)^4/Z^3$.
% \\ 
% \indent
%
%
%
\begin{table}
\caption{\label{T3_IIC}Interelectronic-interaction contributions to the $g$ factor of $^{28}$Si$^{11+}$, $^{132}$Xe$^{51+}$, and $^{208}$Pb$^{79+}$ ions obtained in the Coulomb and different screening potentials: core-Hartree (CH), Kohn-Sham (KS), Dirac-Hartree (DH), and Dirac-Slater (DS) (all without the Latter correction), in units of $10^{-6}$.}
\label{table:var_pot1}
\begin{center}
\vspace{-2.5em}
\begin{tabular}{c S S S S S}
\hline\hline
 &\multicolumn{1}{c}{Coulomb} &\multicolumn{1}{c}{CH} &\multicolumn{1}{c}{KS} &\multicolumn{1}{c}{DH} &\multicolumn{1}{c}{DS}\\
\hline
\multicolumn{6}{l}{$Z = 14$}\\
$\Delta g^{(0)}_{\rm int}$ 
                    &                & 348.2661    & 341.4434        & 417.1982       & 302.7096     \\
$\dgint^{(1)}$      & 321.5903       & -33.5491    & -25.1729        &-104.3024       &  14.6672     \\
$\dgint^{(2)}$      &  -6.8782(1)    &   0.1362(1) &  -1.4929(1)     &   2.0986(1)    &  -2.6630(1)  \\
                    &  -6.8787(1)$^a$ \\                
$\Delta g^{(3+)}_\text{int,L}$
                    & 0.0934(21)     &  -0.0443(10)& 0.0330(22)  &  -0.1833(22)    &    0.0956(23)  \\
                    
$\Delta g^{(3+)}_\text{int,H}$
                    & 0.0000(74)     &   0.0000(14)&  0.0000(18)    &  0.0000(5)   &     0.0000(25)  \\
                    & 0.0000(14)$^a$ \\
                    
Total               & 314.8055(77)   & 314.8089(17)&  314.8106(28)  & 314.8111(23) &  314.8094(34) \\
                    & 314.8058(15)$^a$ \\
\hline                    
\multicolumn{6}{l}{$Z = 54$}\\
$\Delta g^{(0)}_{\rm int}$ 
                    &                &1463.1608    &1414.1410       &1763.0709      &1238.6240     \\
$\dgint^{(1)}$      & 1306.2170      &-164.6779    &-113.4828       &-466.1252      &  63.4535     \\
$\dgint^{(2)}$      &   -7.6565(3)   &   0.1376(2) &  -2.0667(1)    &   1.7041(1)   &  -3.5117(1)  \\
                    &   -7.6569(5)$^a$ \\
$\Delta g^{(3+)}_\text{int,L}$
                    &    0.0308(22)  &  -0.0106(19)&   0.0174(10)    &   -0.0354(17)  &   0.0417(9) \\
$\Delta g^{(3+)}_\text{int,H}$
                    &    0.0000(255) &   0.0000(74)&   0.0000(81)   &   0.0000(41) &    0.0000(101) \\
                    
Total               &  1298.5913(256) &  1298.6099(76)  &  1298.6089(82)   &  1298.6144(44)   &  1298.6075(101) \\
\hline                    
\multicolumn{6}{l}{$Z = 82$}\\
$\Delta g^{(0)}_{\rm int}$ 
                    &                &2448.8325     &2341.3605     &2952.7350      & 2034.2466      \\
$\dgint^{(1)}$      & 2148.2959      &-309.3425     &-198.8050     &-814.6876      &  110.2598      \\
$\dgint^{(2)}$      &   -8.9897(5)   &  -0.1361(1)  &  -3.2341(1)  &   1.3279(1)   &   -5.2159(1)   \\
                    &   -8.9905(6)$^a$ \\
$\Delta g^{(3+)}_\text{int,L}$
                    &    0.0459(24)  &  -0.0057(31) &   0.0232(27) &   -0.0237(23)  &   0.0564(14)  \\
$\Delta g^{(3+)}_\text{int,H}$
                    &    0.0000(283) &   0.0000(122) &   0.0000(115) &   0.0000(86) &    0.0000(130) \\
                    
Total               &  2139.3521(284)   &  2139.3482(126)  &  2139.3446(118)   &  2139.3516(89)     &  2139.3469(131) \\
\hline\hline
\end{tabular} \\
$^a$ Yerokhin {\it et al.} (2021) \cite{yerokhin:2021:022814}
\end{center}
\end{table}
\begin{table}
\caption{\label{T4_res}Interelectronic-interaction contributions to the $g$ factor of lithiumlike ions with $Z= 14$ -- $82$ obtained for the core-Hartree (CH) potential, in units of $10^{-6}$.}
\label{table:final_results}
\centering
\tabcolsep5pt
\begin{tabular}{c S S S S S S S}
\hline\hline
$Z$ 
&\multicolumn{1}{c}{$\Delta g^{(0)}_{\rm int}$} 
&\multicolumn{1}{c}{$\Delta g^{(1)}_{\rm int}$} 
&\multicolumn{1}{c}{$\Delta g^{(2)}_{\rm int}$} 
&\multicolumn{1}{c}{$\Delta g^{(3+)}_\text{int,L}$}
&\multicolumn{1}{c}{$\Delta g^{(3+)}_\text{int,H}$} 
&\multicolumn{1}{c}{Total}          \\
\hline
14  & 348.2661    &  -33.5491    & 0.1362(1)  & -0.0443(10)   & 0.0000(14)    & 314.8089(17)     \\
    & 348.267$^a$ &  -33.549$^a$ & 0.137$^a$  & -0.046(6)$^a$ &               & 314.809(6)$^a$   \\
    &             &              &            &               &               & 314.8058(15)$^b$ \\[1mm]
18  & 452.6955    &  -45.1729    & 0.1301(1)  & -0.0318(12)   & 0.0000(20)    & 407.6209(23)     \\[1mm] 
20  & 505.2339    &  -51.0429    & 0.1291(1)  & -0.0300(8)    & 0.0000(24)    & 454.2902(25)     \\
    & 505.234$^a$ &  -51.042$^a$ & 0.129$^a$  & -0.031(9)$^a$ &               & 454.290(9)$^a$   \\
    &             &              &            &               &               & 454.2834(25)$^b$ \\[1mm]
24  & 611.0988    &  -62.9331    & 0.1300(1)  & -0.0239(17)   & 0.0000(29)    & 548.2718(34)     \\[1mm]
32  & 826.7992    &  -87.5079    & 0.1371(1)  & -0.0185(13)   & 0.0000(40)    & 739.4099(42)     \\[1mm]
40  & 1049.4869   & -113.5613    & 0.1451(1)  & -0.0147(13)   & 0.0000(57)    & 936.0560(58)     \\[1mm]
54  & 1463.1608   & -164.6779    & 0.1376(2)  & -0.0106(19)   & 0.0000(74)    & 1298.6099(76)    \\[1mm]
70  & 1991.9926   & -237.5105    & 0.0466(1)  & -0.0090(24)   & 0.0000(100)   & 1754.520(10)     \\[1mm]
82  & 2448.8325   & -309.3425    &-0.1361(1)  & -0.0057(31)   & 0.0000(122)   & 2139.348(13)     \\
    & 2448.833$^a$& -309.340$^a$ &-0.134$^a$  & -0.015(10)$^a$& 0.000(40)$^a$ & 2139.34(4)$^a$   \\
\hline
\hline
\end{tabular} \\
$^a$ Volotka {\it et al.} (2014) \cite{volotka:2014:253004};
$^b$ Yerokhin {\it et al.} (2021) \cite{yerokhin:2021:022814}.
\end{table}

% ============================= CONCLUSION ==============================
%
\section{CONCLUSION}
In conclusion, the electron correlation effects on the $g$ factor of lithiumlike ions in the range $Z=14-82$ are evaluated with an uncertainty on the level of $10^{-6}$. The first- and second-order interelectronic-interaction corrections are calculated within the rigorous bound-state QED approach, i.e., to all orders in $\alpha Z$. The third- and higher-order contributions are taken into account within the Breit approximation using the recursive perturbation theory. In comparison to previous theoretical calculations, the accuracy of the interelectronic-interaction contributions to the bound-electron $g$ factor in lithiumlike ions is substantially improved.
%

%cbbcb
% ======================== ACKNOWLEDGEMENTS ============================
%
\section*{Acknowledgments}
We thank V. A. Yerokhin for valuable discussions. 
% The QED calculations in Secs.~\ref{subsec:A} and \ref{subsec:B} were supported by the Foundation for the Advancement of Theoretical Physics and Mathematics ``BASIS'', while the part devoted to the Breit approximation, Sec.~\ref{subsec:C}, was supported by the Russian Science Foundation (Grant No. 22-12-00258).
The work was supported by the Russian Science Foundation (Grant No. 22-12-00258).

% ======================== APPENDIX ============================
\section*{APPENDIX: QED formulas for two-photon-exchange contribution}
In this Appendix we present the explicit formulas for the two-photon-exchange contribution to the $g$ factor of lithiumlike ions derived within the two-time Green's function method~\cite{shabaev:2002:119}. We separate the irreducible contributions of the two- and three-electron diagrams, the reducible contribution, including the non-diagrammatic terms, and the counterterm contribution, see Eq.~\eqref{eq:2ph}.
% \section*{APPENDIX A: Two-photon exchange three-electron contribution $\Delta g^{(2)}_{\rm 3el}$}
\subsection*{Three-electron contribution}
\label{sec:A:A}
The three-electron contribution to the two-photon-exchange correction, see Fig. \ref{fig-3el}, is given by the sum, according to the types of diagrams in Fig.~\ref{fig-3el},
\begin{equation}
\Delta g^{(2)}_{\rm 3el} = \Delta g^{(2)}_{\rm 3el, A} + \Delta g^{(2)}_{\rm 3el, B} + \Delta g^{(2)}_{\rm 3el, C} + \Delta g^{(2)}_{\rm 3el, D}\,,
\end{equation}
where
\be
\label{Atype}
\Delta g^{(2)}_{\rm 3el, A} = \frac{1}{m_a} \sum_{b_1,b_2} \sum_{P,Q} (-1)^{P+Q} {\sum_n}'
 \frac{\la Pa Pb_1 | I(\Delta_{Pa  Qa}  ) | \xi_{Qa} n    \ra
       \la  n Pb_2 | I(\Delta_{Pb_2Qb_2}) | Qb_1     Qb_2 \ra}
      {\veps_{Pa} + \veps_{Pb_1} - \veps_{Qa} - \veps_n}\,,
\ee
\be
\label{Btype}
\Delta g^{(2)}_{\rm 3el, B} = \frac{1}{m_a} \sum_{b_1,b_2} \sum_{P,Q} (-1)^{P+Q} {\sum_n}'
 \frac{\la \xi_{Pa} Pb_1 | I(\Delta_{Pa  Qa}  ) | Qa   n    \ra
       \la n        Pb_2 | I(\Delta_{Pb_2Qb_2}) | Qb_1 Qb_2 \ra}
      {\veps_{Pa} + \veps_{Pb_1} - \veps_{Qa} - \veps_n}\,,
\ee
\be
\label{Ctype}
\Delta g^{(2)}_{\rm 3el, C} &=& \frac{1}{2m_a} \sum_{b_1,b_2} \sum_{P,Q} (-1)^{P+Q}\nonumber\\
 &\times&{\sum_{n_1,n_2}}'
 \frac{\la Pa  Pb_1 | I(\Delta_{Pa  Qa  }) | Qa   n_1  \ra
       \la n_1 | U | n_2 \ra
       \la n_2 Pb_2 | I(\Delta_{Pb_2Qb_2}) | Qb_1 Qb_2 \ra}
      {(\veps_{Pa}   + \veps_{Pb_1} - \veps_{Qa}   - \veps_{n_1})
       (\veps_{Qb_1} + \veps_{Qb_2} - \veps_{Pb_2} - \veps_{n_2})}\,,
\ee
\be
\label{Dtype}
\Delta g^{(2)}_{\rm 3el, D} = \frac{1}{m_a} \sum_{b_1,b_2} \sum_{P,Q} (-1)^{P+Q} {\sum_n}'
 \frac{\la Pa \xi_{Pb_1} | I(\Delta_{Pa  Qa}  ) | Qa   n    \ra
       \la n  Pb_2       | I(\Delta_{Pb_2Qb_2}) | Qb_1 Qb_2 \ra}
      {\veps_{Pa} + \veps_{Pb_1} - \veps_{Qa} - \veps_n}\,,
\ee
where
\be
\label{eq:xsi}
|\xi_c \ra = {\sum_n}' \frac{| n \ra \la n | U | c \ra}{\veps_c - \veps_n}\,,
\ee
the prime over the sums means that terms with vanishing denominators should be omitted in the summation, $P$ and $Q$ are permutation operators, which determine the sign $(-1)^{P+Q}$, $\Delta_{PaQb} = \veps_{Pa} - \veps_{Qb}$.
%The sum of terms $\Delta g^{(2)}_{\rm 3el, A}$, $\Delta g^{(2)}_{\rm 3el, B}$, and $\Delta g^{(2)}_{\rm 3el, D}$ corresponds to the contribution $\Delta g^{\text{3el}}_{\text{ir,pwf}}$, Eq.~(43) of Ref.~\cite{yerokhin:2021:022814}, while the term $\Delta g^{(2)}_{\rm 3el, C}$ stands for the part of the contribution $\Delta g^{\text{3el}}_{\text{ir,ver}}$ with $\Xi X/(\Delta_1 \Delta_2) = X/(\Delta_1 \Delta_2)$, Eqs.~(45), (46) of Ref.~\cite{yerokhin:2021:022814}.
%
%%%%%%%%%%%%%%%%%%%%%%%%%%%%%%%%%%%%%%%%%%%%%%%%%%%%%%%%%%%%%%%%%%%
% \section*{APPENDIX B: Two-photon exchange two-electron contribution}
\subsection*{Two-electron contribution}
%%%%%%%%%%%%%%%%%%%%%%%%%%%%%%%%%%%%%%%%%%%%%%%%%%%%%%%%%%%%%%%%%%%
\label{sec:A:B}
The irreducible parts of the two-electron diagrams depicted in Fig. \ref{fig-2el} yield
\begin{equation}
\label{eqA:2ph_2el}
\Delta g^{(2)}_{\rm 2el} = \Delta g^{(2)}_{\rm 2el, lad-W} + \Delta g^{(2)}_{\rm 2el, lad-S} + \Delta g^{(2)}_{\rm 2el, cr-W} + \Delta g^{(2)}_{\rm 2el, cr-S}\,,
\end{equation}
with
\be
\label{ladWtype}
\Delta g^{(2)}_{\rm 2el, lad-W} &=& \frac{1}{m_a} \sum_b \sum_{P,Q} (-1)^{P+Q}
 \frac{i}{\pi} \int_{-\infty}^\infty \rmd\omega  
\nonumber\\ 
 &\times&{\sum_{n_1,n_2}}'\, 
 \frac{\la Pa Pb | I(\omega) | n_1 n_2 \ra
       \la n_1 n_2 | I(\omega+\Delta_{PaQa}) | \xi_{Qa} Qb \ra}
      {(\veps_{Pa}+\omega-u\veps_{n_1})(\veps_{Qb}-\omega-\Delta_{PaQa}-u\veps_{n_2})}\,,
\ee
\be
\label{ladStype}
\Delta g^{(2)}_{\rm 2el, lad-S} &=& \frac{1}{m_a} \sum_b \sum_{P,Q} (-1)^{P+Q}
 \frac{i}{2\pi} \int_{-\infty}^\infty \rmd \omega
 \nonumber \\
 &\times&{\sum_{n_1,n_2,n_3}}'\,
 \frac{\la Pa Pb | I(\omega) | n_1 n_2 \ra \la n_2 | U | n_3 \ra
       \la n_1 n_3 | I(\omega+\Delta_{PaQa}) | Qa Qb \ra}
      {(\veps_{Pa}+\omega-u\veps_{n_1})(\veps_{Qb}-\omega-\Delta_{PaQa}-u\veps_{n_2})
       (\veps_{Qb}-\omega-\Delta_{PaQa}-u\veps_{n_3})}\,,
\ee
\be
\label{crWtype}
\Delta g^{(2)}_{\rm 2el, cr-W} &=& \frac{1}{m_a} \sum_b \sum_{P,Q} (-1)^{P+Q}
 \frac{i}{\pi} \int_{-\infty}^\infty \rmd\omega
\nonumber\\ 
 &\times& {\sum_{n_1,n_2}}'\,
 \frac{\la Pa n_2 | I(\omega) | n_1 Qb \ra
       \la \xi_{Pb} n_1 | I(\omega-\Delta_{PaQa}) | n_2 Qa \ra}
      {(\veps_{Pa}-\omega-u\veps_{n_1})(\veps_{Qb}-\omega-u\veps_{n_2})}\,,
\ee
\be
\label{crStype}
\Delta g^{(2)}_{\rm 2el, cr-S} &=&  \frac{1}{m_a} \sum_b \sum_{P,Q} (-1)^{P+Q}
 \frac{i}{2\pi} \int_{-\infty}^\infty \rmd\omega 
 \nonumber \\
 &\times& {\sum_{n_1,n_2,n_3}}'\,
 \frac{\la Pa n_2 | I(\omega) | n_1 Qb \ra \la n_3 | U | n_2 \ra
       \la Pb n_1 | I(\omega-\Delta_{PaQa}) | n_3 Qa \ra}
      {(\veps_{Pa}-\omega-u\veps_{n_1})(\veps_{Qb}-\omega-u\veps_{n_2})
       (\veps_{Qb}-\omega-u\veps_{n_3})}\,,
\ee
where the prime on the sums indicates that in the summation we omit the
reducible and infrared-divergent terms, namely, those with $\veps_{n_1}+\veps_{n_2}=\veps_a+\veps_b$ in the ladder-W diagrams,
with $\veps_{n_1}=\veps_{Pa},\,\veps_{n_2}=\veps_{Qb}$
in the direct parts of the cross-W diagrams and
$\veps_{n_1}=\veps_{n_2}=\veps_a,\veps_b$
in the exchange parts of the cross-W diagrams,
with $\veps_{n_1}+\veps_{n_2}=\veps_a+\veps_b$,
     $\veps_{n_1}+\veps_{n_3}=\veps_a+\veps_b$, and
     $\veps_{n_2}=\veps_{n_3}=\veps_{Qb}-\Delta_{PaQa}$
in the ladder-S diagrams,
with $\veps_{n_1}=\veps_{Pa},\,\veps_{n_2}=\veps_{Qb}$,
     $\veps_{n_1}=\veps_{Pa},\,\veps_{n_3}=\veps_{Qb}$, and
     $\veps_{n_2}=\veps_{n_3}=\veps_{Qb}$
in the direct parts of the cross-S diagrams,
with $\veps_{n_1}=\veps_{n_2}=\veps_a,\veps_b$,
     $\veps_{n_1}=\veps_{n_3}=\veps_a,\veps_b$, and
     $\veps_{n_2}=\veps_{n_3}=\veps_a,\veps_b$
in the exchange parts of the cross-S diagrams. $u=1-i0$ preserves the proper treatment of poles of the electron propagators.
%These expressions corresponds to the formulas (21) and (25) of Ref.~\cite{yerokhin:2021:022814} but only with the wave function perturbation terms $\delta F_\text{lad,dir}(\omega, n_1 n_2)$, $\delta F_\text{cr,dir}(\omega, n_1 n_2)$ and $\delta F_\text{lad,ex}(\omega, n_1 n_2)$, $\delta F_\text{cr,ex}(\omega, n_1 n_2)$, respectively. Moreover, the last terms in $\delta F_\text{lad,ex}(\omega, n_1 n_2)$ and $\delta F_\text{cr,ex}(\omega, n_1 n_2)$ given by expressions (26) and (27) of Ref.~\cite{yerokhin:2021:022814} are omitted here, since we include them into reducible part.
%
%%%%%%%%%%%%%%%%%%%%%%%%%%%%%%%%%%%%%%%%%%%%%%%%%%%%%%%%%%%%%%%%%%%
% \section*{APPENDIX C: Two-photon exchange reducible contribution}
\subsection*{Reducible contribution}
%%%%%%%%%%%%%%%%%%%%%%%%%%%%%%%%%%%%%%%%%%%%%%%%%%%%%%%%%%%%%%%%%%%
\label{sec:A:C}
The reducible parts of the two-electron diagrams are given by the following expressions,
\be
\Delta g^{(2)}_{\rm red} = \Delta g^{(2)}_{\rm red, E} + \Delta g^{(2)}_{\rm red, F} + \Delta g^{(2)}_{\rm red, G} + \Delta g^{(2)}_{\rm red, 2el}\,,
\ee
%
%%%%%%%%%%%%%%%%%%%%%%%%%%%%%%%%%%%%%%%%%%%%%%%%%%%%%%%%%%%%%%%%%%%
%
where red,E term is given by
\be
\Delta g^{(2)}_{\rm red, E} = \Delta g^{(2)}_{\rm red, Ea} + \Delta g^{(2)}_{\rm red, Eb}
\ee
with
\be
\Delta g^{(2)}_{\rm red, Ea} &=& \frac{1}{m_a} \sum_{b_1, b_2} \sum_{P,Q} (-1)^{P+Q} \la a | U | a \ra {\sum_n}'
\nonumber\\
&\times&\left\{
   \frac12\frac{\la Qb_1 Qb_2 |I(\Delta_{Qb_1a}) | a n \ra \la n a |I(\Delta_{a Pb_2}) | Pb_1 Pb_2 \ra}{(2\veps_b - \veps_a - \veps_n)^2}\right.
 - \frac{\la Qb_1 Qb_2 |I'(\Delta_{Qb_1a}) | a n \ra \la n a |I(\Delta_{a Pb_2}) | Pb_1 Pb_2 \ra}{2\veps_b - \veps_a - \veps_n}
\nonumber\\
&+&\frac{\la Qb_1 Qa |I(\Delta_{Qb_1b_2}) | b_2 n \ra \la n b_2 |I(\Delta_{b_2 Pb_1}) | Pa Pb_1 \ra}{(\veps_a - \veps_n)^2}
 - 2\frac{\la Qb_1 Qa |I'(\Delta_{Qb_1b_2}) | b_2 n \ra \la n b_2 |I(\Delta_{b_2 Pb_1}) | Pa Pb_1 \ra}{\veps_a - \veps_n}
\nonumber\\
&+&\frac{\la Qa Qb_1 |I(\Delta_{Qab_1}) | b_1 n \ra \la n b_2 |I(\Delta_{b_2 Pb_2}) | Pa Pb_2 \ra}{(\veps_a - \veps_n)^2}
 - 2\frac{\la Qa Qb_1 | I'(\Delta_{Qab_1}) | b_1 n \ra \la n b_2 |I(\Delta_{b_2 Pb_2}) | Pa Pb_2 \ra }{\veps_a - \veps_n}
\nonumber\\
&-&\left. 2\frac{\la  Qa Qb_2|I'(\Delta_{Qaa}) | a n \ra \la n b_1 |I(\Delta_{b_1 Pb_1}) | Pb_2 Pb_1 \ra}{\veps_b - \veps_n}\right\}
\ee
and
\be
\Delta g^{(2)}_{\rm red, Eb} &=& \frac{1}{m_a} \sum_{b_1, b_2}\sum_{P,Q} (-1)^{P+Q} \la b_2 | U | b_2 \ra {\sum_n}' \left\{
   \frac{\la Qb_2 Qa |I(\Delta_{Qb_2b_1})| b_1 n \ra \la n b_1 |I(\Delta_{b_1 Pb_2})| Pa Pb_2 \ra}{(\veps_a - \veps_n)^2}\right.\nonumber\\
&+&\frac{\la Qb_2 Qb_1 |I(\Delta_{Qb_2a})| a n \ra \la n a |I(\Delta_{a Pb_2})| Pb_1 Pb_2\ra}{(2\veps_b - \veps_a - \veps_n)^2}
 - \frac{\la Qb_2 Qb_1 |I'(\Delta_{Qb_2a})| a n \ra \la n a |I(\Delta_{a Pb_2})| Pb_1 Pb_2 \ra}{2\veps_b - \veps_a - \veps_n}
\nonumber \\
&+&2\frac{\la Qb_2 Qb_1 |I(0)| b_1 n \ra \la n a |I(\Delta_{a Pa})| Pb_2 Pa \ra}{(\veps_b - \veps_n)^2}
 + 2\frac{\la Qb_2 Qb_1 |I(0)| b_1 n \ra \la n a |I'(\Delta_{a Pa})| Pb_2 Pa \ra}{\veps_b - \veps_n}
\nonumber \\
&+&\frac{\la Qa Qb_1 |I(\Delta_{Qab_2}) | b_2 n \ra \la n b_2 |I(\Delta_{b_2 Pb_1})| Pa Pb_1 \ra}{(\veps_a - \veps_n)^2}
 - 2\frac{\la Qa Qb_1 |I'(\Delta_{Qab_2})| b_2 n \ra \la n b_2 |I(\Delta_{b_2 Pb_1})| Pa Pb_1 \ra}{\veps_{a} - \veps_n}
\nonumber \\
&-&\left. 2\frac{\la Qb_2 Qa |I'(\Delta_{Qb_2 b_2})| b_2 n \ra \la n b_1 |I(\Delta_{b_1 Pb_1}) | Pa Pb_1 \ra}{\veps_a - \veps_n}\right\}\,.
\ee
%
%%%%%%%%%%%%%%%%%%%%%%%%%%%%%%%%%%%%%%%%%%%%%%%%%%%%%%%%%%%%%%%%%%
%
The term red,F can be written as
\be
\Delta g^{(2)}_{\rm red, F} = \Delta g^{(2)}_{\rm red, Fa} + \Delta g^{(2)}_{\rm red, Fb}
\ee
with
\be
\Delta g^{(2)}_{\rm red, Fa} &=& \frac{2}{m_a} \sum_{b_1, b_2} \sum_{P,Q} (-1)^{P+Q}
\sum_n^{\veps_n = \veps_a}
\left\{
\left[
   \la \xi_{Pb_2} Pa | I'(\Delta_{Pb_2b_2}) | b_2 n \ra
 + \la Pb_2 \xi_{Pa} | I'(\Delta_{Pb_2b_2}) | b_2 n \ra
\right.
\right.
\nonumber\\
&-&\left.
    \la Pb_2 P\xi'_a | I(\Delta_{Pb_2b_2}) | b_2 n \ra
    \right]
    \la n b_1 | I(\Delta_{b_1 Qb_1}) | Qa Qb_1 \ra
 +  \left[
    \la \xi_{Pb_2} Pa | I(\Delta_{Pb_2b_2}) | b_2 n \ra
    \right.
\nonumber\\
&+&\left.
   \la Pb_2 \xi_{Pa} | I(\Delta_{Pb_2b_2}) | b_2 n \ra
   \right]
   \la n b_1 | I'(\Delta_{b_1 Qb_1}) | Qa Qb_1 \ra
 - \la Pb_1 Pa | I'(\Delta_{Pb_1b_2}) | \xi_{b_2} n \ra
\nonumber\\
&\times&
   \la n b_2 | I(\Delta_{b_2 Qb_1}) | Qa Qb_1 \ra
 + \la Pb_1 \xi_{Pa} | I(\Delta_{Pb_1b_2}) | b_2 n \ra
   \la n b_2 | I'(\Delta_{b_2Qb_1}) | Qa Qb_1 \ra
\nonumber\\
&+&\sum_m'
   \left[
   \frac{\la Pb_1 Pa | I(\Delta_{Pb_1b_2}) | b_2 m \ra \la m | U | n \ra \la n b_2 | I(\Delta_{b_2Qb_1}) | Qa Qb_1 \ra}{(\veps_a - \veps_m)^2}
   \right.
\nonumber\\
&-&\frac{\la Pb_1 Pa | I'(\Delta_{Pb_1b_2}) | b_2 m \ra \la m | U | n \ra \la n b_2 | I(\Delta_{b_2Qb_1}) | Qa Qb_1 \ra}{\veps_a - \veps_m}
\nonumber\\
&+&\left.\left.
   \frac{\la Pb_1 Pa | I(\Delta_{Pb_1b_2}) | b_2 m \ra \la m | U | n \ra \la n b_2 | I'(\Delta_{b_2Qb_1}) | Qa Qb_1 \ra}{\veps_a - \veps_m}
   \right]\right\}
\nonumber\\
&-&\frac{2}{m_a} \sum_{b_1, b_2}\sum_n^{\veps_n = \veps_a}
   \left\{
   \left[
   \la \xi_a b_1 | I'(\Delta_{ab}) | b_2 n \ra
 + \la a \xi_{b_1} | I'(\Delta_{ab}) | b_2 n \ra
   \right]
   \la n b_2 | I(\Delta_{ab}) | b_1 a \ra
   \right.
 + \la a b_1 | I(\Delta_{ab}) | \xi'_{b_2} n \ra
\nonumber\\
&\times&
   \la n b_2 | I(\Delta_{ab}) | b_1 a \ra
 + \left[
   \la \xi_a b_1 | I(\Delta_{ab}) | b_2 n \ra
 + \la a b_1 | I(\Delta_{ab}) | \xi_{b_2} n \ra
   \right]
   \la n b_2 | I'(\Delta_{ab}) | b_1 a \ra
\nonumber\\
&+&\left.
   \sum_m'
   \frac{\la b_1 a | I(0) | b_2 m \ra \la m | U | n \ra \la n b_2 | I(0) | a b_1 \ra}{(\veps_a - \veps_m)^2}
   \right\}
\ee
and
\be
\Delta g^{(2)}_{\rm red, Fb} &=& \frac{2}{m_a} \sum_{b_1, b_2} \sum_{P,Q} (-1)^{P+Q}
\sum_n^{\veps_n = \veps_b}
\left\{
\left[
   \la \xi_{Pa} Pb_2 | I'(\Delta_{Paa}) | a n \ra 
 + \la Pa \xi_{Pb_2} | I'(\Delta_{Paa}) | a n \ra
\right.
\right.
\nonumber\\
&-&\left.
   \la Pa P\xi'_{b_2} | I(\Delta_{Paa}) | a n \ra
   \right]
   \la n b_1 | I(0) | Qb_2 Qb_1 \ra
 - \la Pb_2 P\xi'_{b_1} | I(0) | b_2 n \ra \la n a | I(\Delta_{a Qa}) | Qb_1 Qa \ra
\nonumber\\
&-&\left.
   \left[
   \la Pb_2 \xi_{Pb_1} | I(0) | b_2 n \ra
 + \la \xi_{Pb_2} Pb_1 | I(0) | b_2 n \ra
   \right]
   \la n a | I'(\Delta_{aQa}) | Qb_1 Qa \ra
   \right\}\,,
\ee
where $| \xi'_c \ra = \partial/\partial\veps_c\,| \xi_c \ra$.
%
%%%%%%%%%%%%%%%%%%%%%%%%%%%%%%%%%%%%%%%%%%%%%%%%%%%%%%%%%%%%%%%%%%
%
The term red,G can be expressed by 
\be
\Delta g^{(2)}_{\rm red, G} = \Delta g^{(2)}_{\rm red, Ga} + \Delta g^{(2)}_{\rm red, Gb}
\ee
with
\be
\Delta g^{(2)}_{\rm red, Ga} &=& \frac{1}{m_a} \sum_{b_1, b_2}\sum_P(-1)^P\sum_n^{\veps_n=\veps_a}
\left\{
   \la a | U | a \ra
   \left[
   \la a b_1 | I''(\Delta_{ab}) | b_2 n \ra \la n b_2 | I(\Delta_{b_2Pb_1}) | Pa Pb_1 \ra
\right.
\right.
\nonumber \\
&-&\la a b_1 | I'(\Delta_{ab}) | b_2 n \ra \la n b_2 | I'(\Delta_{b_2Pb_1}) | Pa Pb_1 \ra
 - \la a b_1 | I''(\Delta_{ab}) | b_1 n \ra \la n b_2 | I(\Delta_{b_2Pb_2}) | Pa Pb_2 \ra
\nonumber \\
&+&\left.
   \la a b_1 | I'(\Delta_{ab}) | b_1 n \ra \la n b_2 | I'(\Delta_{b_2Pb_2}) | Pa Pb_2 \ra
   \right]
 + \la n | U | n \ra
   \left[
   \la a b_1 | I''(\Delta_{ab}) | b_2 n \ra
   \right.
\nonumber\\
&\times&\left.\left.
   \la n b_2 | I(\Delta_{b_2Pb_1}) | Pa Pb_1 \ra
 - \la a b_1 | I'(\Delta_{ab}) | b_2 n \ra
   \la n b_2 | I'(\Delta_{b_2Pb_1}) | Pa Pb_1 \ra
\right]\right\}
\nonumber\\
&+&\frac{1}{m_a} \sum_{b_1, b_2}\sum_P(-1)^P\sum_n^{\veps_n=\veps_b}
   \la a | U | a \ra
   \la b_1 a | I''(\Delta_{ab}) | a n \la \ra n b_2 | I(0) | Pb_1 Pb_2 \ra
\ee
and
\be
\Delta g^{(2)}_{\rm red, Gb} &=& \frac12\frac{1}{m_a} \sum_{b_1, b_2}\sum_n^{\veps_n=\veps_a} \la b_2 | U | b_2 \ra
   \left\{
   2 \la a b_1 | I''(\Delta_{ab}) | b_2 n \ra \la n b_2 | I(\Delta_{ab}) | b_1 a \ra
   \right.
\nonumber\\
&+&2 \la a b_1 | I'(\Delta_{ab}) | b_2 n \ra \la n b_2 | I'(\Delta_{ab}) | b_1 a \ra
 -   \la b_1 a | I(0) | b_2 n \ra \la n b_2 | I''(\Delta_{ab}) | b_1 a \ra
\nonumber\\
&+&2 \la a b_2 | I''(\Delta_{ab}) | b_1 n \ra \la n b_1 | I(\Delta_{ab}) | b_2 a \ra
 + 2 \la a b_2 | I'(\Delta_{ab}) | b_1 n \ra \la n b_1 | I'(\Delta_{ab}) | b_2 a \ra
\nonumber\\
&-&  \la b_2 a | I(0) | b_1 n \ra \la n b_1 | I''(\Delta_{ab}) | b_2 a \ra
 - 2 \la a b_2 | I''(\Delta_{ab}) | b_2 n \ra \la n b_1 | I(0) | a b_1 \ra
\nonumber\\
&+&\left.
   2 \la a b_2 | I''(\Delta_{ab}) | b_2 n \ra \la n b_1 | I(\Delta_{ab}) | b_1 a \ra
 - 2 \la a b_2 | I'(\Delta_{ab}) | b_2 n \ra \la n b_1 | I'(\Delta_{ab}) | b_1 a \ra
   \right\}
\nonumber\\
&+&\frac{1}{m_a} \sum_{b_1, b_2}\sum_P(-1)^P\sum_n^{\veps_n=\veps_b} \la b_2 | U | b_2 \ra
   \left\{
   \la b_1 b_2 | I''(0) | b_2 n \ra \la n a | I(\Delta_{aPa}) | Pb_1 Pa \ra
   \right.
\nonumber\\
&-&\left.
   \la b_2 b_1 | I''(0) | b_1 n \ra \la n a | I(\Delta_{aPa}) | Pb_2 Pa \ra
 - \la b_2 a | I''(\Delta_{ab}) | a n \ra \la n b_1 | I(0) | Pb_2 Pb_1 \ra
   \right\}\,.
\ee
%
%%%%%%%%%%%%%%%%%%%%%%%%%%%%%%%%%%%%%%%%%%%%%%%%%%%%%%%%%%%%%%%%%%%%%%%%%%%%%%%%
%%%%%%%%%%%%%%%%%%%%%%%%%%%%%%%%%%%%%%%%%%%%%%%%%%%%%%%%%%%%%%%%%%%%%%%%%%%%%%%%
%
The reducible two-electron term is found to be 
\be
\Delta g^{(2)}_{\rm red, 2el} &=& \Delta g^{(2)}_{\rm red, 2el, lad-W} + \Delta g^{(2)}_{\rm red, 2el, lad-S} + \Delta g^{(2)}_{\rm red, 2el, cr-W} + \Delta g^{(2)}_{\rm red, 2el, cr-S}\,,
\ee
where
\be
\Delta g^{(2)}_{\rm red, 2el, lad-W} &=&
 - \frac{1}{m_a} \sum_b \frac{i}{\pi} \int_{-\infty}^\infty \rmd\omega
\nonumber\\
&\times&
\left\{
\sum_{n_1,n_2}^{\veps_{n_1}+\veps_{n_2} = \veps_a + \veps_b} \left[
\sum_P (-1)^P
\left(
   \frac{\la a b | I(\omega) | n_1 n_2 \ra \la n_1 n_2 | I(\omega+\Delta_{aPa}) | P\xi_a Pb \ra}{(\veps_a + \omega - u\veps_{n_1})^2} 
\right.
\right.
\right.
\nonumber\\
&-&\frac12\frac{\la a b | I(\omega) | n_1 n_2 \ra \la n_1 n_2 | I(\omega+\Delta_{aPa}) | Pa Pb \ra \la a | U | a \ra}{(\veps_a + \omega - u\veps_{n_1})^3}
\nonumber\\
&+&\frac{\la Pa Pb | I(\omega + \Delta_{Paa}) | n_1 n_2 \ra \la n_1 n_2 | I(\omega) | a \xi_b \ra}{(\veps_a - \omega - u\veps_{n_1})^2}
\nonumber\\
&+&
\left.
\frac12\frac{\la Pa Pb | I(\omega + \Delta_{Paa}) | n_1 n_2 \ra \la n_1 n_2 | I(\omega) | a b \ra \la b | U | b \ra}{(\veps_a - \omega - u\veps_{n_1})^3}
\right)
\nonumber\\
&-&\frac12\frac{\la a b | I(\omega) | n_1 n_2 \ra \la n_1 n_2 | I'(\omega+\Delta_{ab}) | b a \ra \la a | U | a \ra}{(\veps_a + \omega - u\veps_{n_1})^2}
\nonumber\\
&-&
\left.
\frac12\frac{\la b a | I'(\omega + \Delta_{ba}) | n_1 n_2 \ra \la n_1 n_2 | I(\omega) | a b \ra \la b | U | b \ra}{(\veps_a - \omega - u\veps_{n_1})^2}
\right]
\nonumber\\
&+&
\frac12
\sum_{n_1,n_2}^{\veps_{n_1}+\veps_{n_2} \neq \veps_a + \veps_b}
\left[
\sum_P (-1)^P
\left(
\frac{\la a b | I(\omega) | n_1 n_2 \ra \la n_1 n_2 | I(\omega+\Delta_{aPa}) | Pa Pb \ra \la a | U | a \ra}{(\veps_a + \omega - u\veps_{n_1})^2 (\veps_b - \omega - u\veps_{n_2})} 
\right.
\right.
\nonumber\\
&+&
\left.
\frac{\la Pa Pb | I(\omega + \Delta_{Paa}) | n_1 n_2 \ra \la n_1 n_2 | I(\omega) | a b \ra \la b | U | b \ra}{(\veps_a - \omega - u\veps_{n_1})(\veps_b + \omega - u\veps_{n_2})^2}
\right)
\nonumber\\
&+&\frac{\la a b | I(\omega) | n_1 n_2 \ra \la n_1 n_2 | I'(\omega+\Delta_{ab}) | b a \ra \la a | U | a \ra}{(\veps_a + \omega - u\veps_{n_1}) (\veps_b - \omega - u\veps_{n_2})} 
\nonumber\\
&+&
\left.
\left.
\frac{\la b a | I(\omega + \Delta_{ba}) | n_1 n_2 \ra \la n_1 n_2 | I'(\omega) | a b \ra \la b | U | b \ra}{(\veps_a - \omega - u\veps_{n_1})(\veps_b + \omega - u\veps_{n_2})}
\right]
\right\}
\nonumber\\
&-&
\frac{1}{2m_a}
\sum_{b,n_1,n_2}^{\veps_{n_1}+\veps_{n_2} = \veps_a + \veps_b}
\la a b | I(\Delta_{an_1}) | n_1 n_2 \ra \la n_1 n_2 | I''(\Delta_{n_1b}) | b a \ra \left(\la a | U| a \ra - \la b | U| b \ra \right),
\ee
%
%%%%%%%%%%%%%%%%%%%%%%%%%%%%%%%%%%%%%%%%%%%%%%%%%%%%%%%%%%%%%%%%%%%%%%%%%%%%%%%%
%
\be
\Delta g^{(2)}_{\rm red, 2el, lad-S} &=&
 - \frac{1}{m_a} \sum_b \sum_P (-1)^P \frac{i}{\pi} \int_{-\infty}^\infty \rmd\omega
\nonumber\\
&\times&
\left\{
   \sum_{n_1,n_2,n_3}^{(i)}
   \frac{\la a b | I(\omega) | n_1 n_2 \ra \la n_2 | U | n_3 \ra \la n_1 n_3 | I(\omega + \Delta_{aPa}) | Pa Pb \ra}{(\veps_a + \omega -u\veps_{n_1})^2(\veps_b - \omega -u\veps_{n_2})}
\right.
\nonumber \\
&+&\sum_{n_1,n_2,n_3}^{(ii)}
   \frac{\la Pb Pa | I(\omega + \Delta_{Pbb}) | n_1 n_2 \ra \la n_2 | U | n_3 \ra \la n_1 n_3 | I(\omega) | b a \ra}{(\veps_b + \omega -u\veps_{n_1})^2(\veps_a - \omega -u\veps_{n_3})}
\nonumber \\
&+&\frac12\sum_{n_1,n_2,n_3}^{(iii)}
   \frac{\la a b | I(\omega) | n_1 n_2 \ra \la n_2 | U | n_3 \ra \la n_1 n_3 | I(\omega + \Delta_{aPa}) | Pa Pb \ra}{(\veps_a + \omega -u\veps_{n_1})^3}
\nonumber \\
&+&
\left.
   \frac12\sum_{n_1,n_2,n_3}^{(iii)}
   \frac{\la Pb Pa | I(\omega + \Delta_{Pbb}) | n_1 n_2 \ra \la n_2 | U | n_3 \ra \la n_1 n_3 | I(\omega) | b a \ra}{(\veps_b + \omega -u\veps_{n_1})^3}
\right\}
\nonumber\\
&+&\frac{1}{m_a} \sum_b \sum_{P,Q} (-1)^{P+Q}
   \frac{i}{2\pi} \int_{-\infty}^\infty \rmd \omega
\nonumber \\
&\times&{\sum_{n_1,n_2,n_3}^{(iv)}}\,
 \frac{\la Pa Pb | I(\omega) | n_1 n_2 \ra \la n_2 | U | n_3 \ra \la n_1 n_3 | I(\omega+\Delta_{PaQa}) | Qa Qb \ra}{(\veps_{Pa}+\omega-u\veps_{n_1})(\veps_{Qb}-\omega-\Delta_{PaQa}-u\veps_{n_2})(\veps_{Qb}-\omega-\Delta_{PaQa}-u\veps_{n_3})}
\nonumber\\
&+&\frac{1}{m_a}\sum_b\sum_{n_1,n_2,n_3}^{\veps_{n_1}=\veps_b\text{\;and\;}\veps_{n_2}=\veps_{n_3}=\veps_a} \la a b | I(\omega) | n_1 n_2 \ra \la n_2 | U | n_3 \ra \la n_1 n_3 | I''(0) | b a \ra\,,
\ee
here, $(i)$ stands for the restrictions $\veps_{n_1} + \veps_{n_3} = \veps_a + \veps_b$ together with $\veps_{n_1} + \veps_{n_2} \neq \veps_a + \veps_b$, $(ii)$ corresponds to the $\veps_{n_1} + \veps_{n_2} = \veps_a + \veps_b$ together with $\veps_{n_1} + \veps_{n_3} \neq \veps_a + \veps_b$, $(iii)$ is shortening of $\veps_{n_1} + \veps_{n_2} = \veps_{n_1} + \veps_{n_3} = \veps_a + \veps_b$, and $(iv)$ stands for $\veps_{n_2} = \veps_{n_2} = \veps_{Qb} - \Delta_{PaQa}$ together with $\veps_{n_1} \neq \veps_{Qa} - \Delta_{PbQb}$,
%
%%%%%%%%%%%%%%%%%%%%%%%%%%%%%%%%%%%%%%%%%%%%%%%%%%%%%%%%%%%%%%%%%%%%%%%%%%%%%%%%
%
\be
\Delta g^{(2)}_{\rm red, 2el, cr-W} &=&
\frac{1}{m_a} \sum_b \sum_{P,Q} (-1)^{P+Q}
\frac{i}{\pi} \int_{-\infty}^\infty \rmd\omega
\sum_{n_1,n_2}^{(i)}
\frac{\la Pa n_2 | I(\omega) | n_1 Qb \ra \la \xi_{Pb} n_1 | I(\omega-\Delta_{PaQa}) | n_2 Qa \ra}{(\veps_{Pa}-\omega-u\veps_{n_1})(\veps_{Qb}-\omega-u\veps_{n_2})}
\nonumber\\
&-&\frac{1}{m_a} \sum_b \frac{i}{\pi} \int_{-\infty}^\infty \rmd\omega \sum_{n_1,n_2}
   \left\{
   \frac{\la a n_2 | I(\omega) | n_1 b \ra \la b n_1 | I(\omega) | n_2 a \ra \la b | U | b \ra}{(\veps_a - \omega - u\veps_{n_1})(\veps_b - \omega - u\veps_{n_2})^2}
   \right.
\nonumber\\
&-&\frac{\la a n_2 | I(\omega) | n_1 a \ra \la b n_1 | I'(\omega+\Delta_{ab}) | n_2 b \ra \la b | U | b \ra}{(\veps_a + \omega - u\veps_{n_1})(\veps_a + \omega - u\veps_{n_2})}
 + \frac{\la b n_2 | I(\omega) | n_1 a \ra \la a n_1 | I(\omega) | n_2 b \ra \la a | U | a \ra}{(\veps_b - \omega - u\veps_{n_1})(\veps_a - \omega - u\veps_{n_2})^2}
\nonumber\\
&-&\left.
   \frac{\la b n_2 | I(\omega) | n_1 b \ra \la a n_1 | I'(\omega-\Delta_{ab}) | n_2 a \ra \la a | U | a \ra}{(\veps_b + \omega - u\veps_{n_1})(\veps_b + \omega - u\veps_{n_2})}
   \right\}
\,,
\ee
here, $(i)$ means $\veps_{n_1} = \veps_{Pa}$ and $\veps_{n_2} = \veps_{Qb}$ in the direct parts and $\veps_{n_1} = \veps_{n_2} = \veps_a$ or $\veps_{n_1} = \veps_{n_2} = \veps_b$ in the exchange parts,
%
%%%%%%%%%%%%%%%%%%%%%%%%%%%%%%%%%%%%%%%%%%%%%%%%%%%%%%%%%%%%%%%%%%%%%%%%%%%%%%%%
%
\be
\Delta g^{(2)}_{\rm red, 2el, cr-S} &=& \frac{1}{m_a} \sum_b \sum_{P,Q} (-1)^{P+Q}
\frac{i}{2\pi} \int_{-\infty}^\infty \rmd\omega 
\nonumber \\
&\times& \sum_{n_1,n_2,n_3}^{(i)}\,
 \frac{\la Pa n_2 | I(\omega) | n_1 Qb \ra \la n_3 | U | n_2 \ra
       \la Pb n_1 | I(\omega-\Delta_{PaQa}) | n_3 Qa \ra}
      {(\veps_{Pa}-\omega-u\veps_{n_1})(\veps_{Qb}-\omega-u\veps_{n_2})
       (\veps_{Qb}-\omega-u\veps_{n_3})}\,,
\ee
here, $(i)$ means the summation over ($\veps_{n_1}=\veps_{Pa}$ and $\veps_{n_2}=\veps_{Qb}$) or ($\veps_{n_1}=\veps_{Pa}$ and $\veps_{n_3}=\veps_{Qb}$) or $\veps_{n_2} = \veps_{n_3} = \veps_{Qb}$ in the direct parts, and over $\veps_{n_1}=\veps_{n_2}=\veps_a$ or $\veps_{n_1}=\veps_{n_2}=\veps_b$ or $\veps_{n_1}=\veps_{n_3}=\veps_a$ or $\veps_{n_1}=\veps_{n_2}=\veps_b$ or
$\veps_{n_2}=\veps_{n_3}=\veps_a$ or $\veps_{n_1}=\veps_{n_2}=\veps_b$ in the exchange parts.
%
%%%%%%%%%%%%%%%%%%%%%%%%%%%%%%%%%%%%%%%%%%%%%%%%%%%%%%%%%%%%%%%%%%%%%%%%%%%%%%%%
% \section*{APPENDIX D: Two-photon exchange counterterm contribution}
\subsection*{Counterterm contribution}
%%%%%%%%%%%%%%%%%%%%%%%%%%%%%%%%%%%%%%%%%%%%%%%%%%%%%%%%%%%%%%%%%%%%%%%%%%%%%%%%
%
%
The formal expressions corresponding to the counterterm diagrams depicted in Fig.~\ref{fig-countertems} are given by
\be
\Delta g^{(2)}_{\rm ct} = \Delta g^{(2)}_{\rm ct-1} + \Delta g^{(2)}_{\rm ct-2}\,,
\ee
where
\be
\label{ct-1}
\Delta g^{(2)}_{\rm ct-1} &=& \frac{2}{m_a} \sum_b \sum_{P,Q} (-1)^{P+Q}\left\{ \sum_n' \left[
   \sum_m'
   \frac{\la Pa | V_{\rm scr} | n \ra \la n | U | m \ra \la m Pb | I(\Delta_{PbQb}) | Qa Qb \ra}{(\veps_{Pa} - \veps_n)(\veps_{Pa} - \veps_m)}
\right.
\right.
\nonumber\\
&+&\frac{\la \xi_{Pa} | V_{\rm scr} | n \ra \la n Pb | I(\Delta_{PbQb}) | Qa Qb \ra}{\veps_{Pa} - \veps_n}
 + \frac{\la \xi_{Pa} Pb | I(\Delta_{PbQb}) | n Qb \ra \la n | V_{\rm scr} | Qa \ra}{\veps_{Qa} - \veps_n}
\nonumber\\
&+&\frac{\la Pa | V_{\rm scr} | n \ra \la n \xi_{Pb} | I(\Delta_{PbQb}) | Qa Qb \ra}{\veps_{Pa} - \veps_n}
 + \frac{\la \xi_{Pa} Pb | I(\Delta_{PbQb}) | Qa n \ra \la n | V_{\rm scr} | Qb \ra}{\veps_{Qb} - \veps_n}
\nonumber\\
&+&\frac{\la Pa | V_{\rm scr} | n \ra \la n Pb | I'(\Delta_{PaQa}) | Qa Qb \ra}{\veps_{Pa} - \veps_n}\left(\la Pa | U | Pa \ra - \la Pb | U | Pb \ra\right)
\nonumber\\
&-&\left.
   \frac{\la Pa | V_{\rm scr} | n \ra \la n Pb | I'(\Delta_{PaQa}) | Qa Qb \ra}{(\veps_{Pa} - \veps_n)^2} \la Pa | U | Pa \ra
   \right]
 + \la Pa Pb | I(\Delta_{PaQa}) | Qa Qb \ra \la \xi'_{Qb} | V_{\rm scr} | Qb \ra
\nonumber\\
&+&\la Pa Pb | I'(\Delta_{PaQa}) | Qa Qb \ra \la \xi_{Qb} | V_{\rm scr} | Qb \ra
 + \la \xi_{Pa} Pb | I'(\Delta_{PaQa}) | Qa Qb \ra
\nonumber\\
&\times&
   \left(\la Pa | V_{\rm scr} | Pa \ra - \la Pb | V_{\rm scr} | Pb \ra\right)
 + \la \xi'_{Pa} Pb | I(\Delta_{PaQa}) | Qa Qb \ra \la Pa | V_{\rm scr} | Pa \ra
\nonumber\\
&+&\left.
   \frac14
   \la Pa Pb | I''(\Delta_{PaQa}) | Qa Qb \ra \left( \la Pa | V_{\rm scr} | Pa \ra - \la Qa | V_{\rm scr} | Qa \ra \right)^2\right\}
\ee
corresponds to the five diagrams from the upper part of Fig.~\ref{fig-countertems}, and
\be
\Delta g^{(2)}_{\rm ct-2} &=& \frac{1}{m_a}\sum_{n,m}'
   \frac{\la a | V_{\rm scr} | n \ra \la n | U | m \ra \la m | V_{\rm scr} | a \ra}{(\veps_a - \veps_n)(\veps_a - \veps_m)}
 + \frac{2}{m_a} \la \xi'_a | V_{\rm scr} | a \ra \la a | V_{\rm scr} | a \ra
\nonumber\\
&+&\frac{2}{m_a}\sum_n'\left[
   \frac{\la \xi_a | V_{\rm scr} | n \ra \la n | V_{\rm scr} | a \ra}{\veps_a - \veps_n}
 - \frac12
   \frac{\la a | V_{\rm scr} | n \ra \la n | V_{\rm scr} | a \ra}{(\veps_a - \veps_n)^2} \la a \vert U\vert a \ra \right]
\ee
stays for the two diagrams from the lower part of Fig.~\ref{fig-countertems}.
%
%=======================================================================
%
% ======================================================================
%
\bibliographystyle{apsrev4-1}
\bibliography{liter}
%
% ======================================================================
\end{document}